%% file: MAIN.tex
\RequirePackage[hyphens]{url}
\documentclass[conference]{IEEEtran}

\usepackage{tikz}
\usepackage{amsmath}
\usepackage{xspace}

\usepackage{filecontents}

\usepackage{array} 
\usepackage{multirow}
\usepackage{comment}
\usepackage{graphicx}
\usepackage{subcaption}
\usepackage{algorithm}
\usepackage{algpseudocode}
\usepackage{amssymb}
\usepackage{todonotes}
\let\labelindent\relax
\usepackage{enumitem}
\PassOptionsToPackage{hyphens}{url}
\newcommand{\BfPara}[1]{{\noindent\bf#1.}}

\newcommand{\etal}{{et al.}\xspace}

\newcommand{\thesystem}{{CELEST}}

\IEEEoverridecommandlockouts
\begin{document}

\date{}

\title{\thesystem: Federated Learning for Globally Coordinated Threat Detection}

\author{
	\IEEEauthorblockN{Talha Ongun$^*$, Simona	Boboila, Alina	Oprea, Tina	Eliassi-Rad}
	\IEEEauthorblockA{Northeastern University}
	\IEEEauthorblockN{Jason	Hiser, Jack	Davidson}
	\IEEEauthorblockA{University of Virginia}
}

\maketitle

\begin{abstract}

The cyber-threat landscape has evolved tremendously in recent years, with new threat variants emerging daily, and large-scale coordinated campaigns becoming more prevalent. 
In this study, we propose \thesystem\  (\underline{C}ollaborativ\underline{E} \underline{LE}arning for \underline{S}calable \underline{T}hreat detection), a federated machine learning framework for global threat detection over HTTP, which is one of the most commonly used protocols for malware dissemination and communication. \thesystem\  leverages federated learning in order to collaboratively train a global model across multiple clients who keep their data locally. Through a novel active learning component integrated with the federated learning technique, our system continuously discovers and learns the behavior of new, evolving, and globally-coordinated cyber threats. We show that \thesystem\ is able to expose attacks that are largely invisible to individual organizations. For instance, in one challenging attack scenario with data exfiltration malware, the global model achieves a three-fold increase in Precision-Recall AUC compared to the local model. 
We also design a poisoning detection and mitigation method, DTrust, specifically designed for federated learning in the collaborative threat detection domain. 
We deploy \thesystem\ on two university networks and show that it is able to detect the malicious HTTP communication with high precision and low false positive rates. Furthermore, during its deployment, \thesystem\ detected a set of previously unknown 42 malicious URLs and 20 malicious domains in one day, which were confirmed to be malicious by VirusTotal.

\end{abstract}

{\footnotesize \textsuperscript{*}Authors are ordered by contribution.}

\input{intro}
\input{problem}

\input{method}

\input{evaluation}

\input{discussion}

\input{related_work}

\input{conclusion}

\section*{Acknowledgements}
We thank Afsah Anwar, Alastair Nottingham, Molly Buchanan, Mark Gardner, Jeffry Lang, and Jeffrey Collyer for their help throughout the project.
This research was sponsored by  contract
number W911NF-18-C0019 with the U.S. Army Contracting Command - Aberdeen Proving Ground (ACC-APG) and the Defense Advanced Research Projects Agency (DARPA),  W911NF-21-10322, and 
by the U.S. Army Combat Capabilities Development Command
Army Research Laboratory under Cooperative Agreement Number
W911NF-13-2-0045 (ARL Cyber Security CRA). The views contained in this document are those of the authors and should not be interpreted as representing the official policies, either expressed or implied, of the ACC-APG, DARPA, Combat Capabilities Development Command Army Research Laboratory or the U.S. Government. The U.S. Government is authorized to reproduce and distribute reprints for Government purposes notwithstanding any copyright notation here on. This project
was also funded by NSF under grant CNS-1717634.

\bibliographystyle{IEEEtranS}
\bibliography{malware}

\input{appendix}

\end{document}

%% file: intro.tex
\section{Introduction}

Modern cyber attacks have become sophisticated, coordinated, and are operating at global scale. We have witnessed globally-coordinated campaigns 
with the ability to spread to hundred of thousands of victim machines on the Internet~\cite{wannacry, antonakakis2017understanding}. While attackers exploit a wide range of vulnerabilities in various protocols, HTTP has become one of the prevalent communication protocols for malware dissemination~\cite{nazca_ndss2014}. To attackers' advantage, malicious communication over the HTTP protocol can easily blend in with the large volumes of benign traffic and is rarely blocked. Existing defenses for HTTP malware include network intrusion detection systems, as well as machine learning (ML) methods applied to domain names~\cite{bilge2014exposure,de2021compromised}, URLs~\cite{ma2009beyond,mamun2016detecting} or HTTP logs~\cite{bartos2016,bortolameotti2017decanter,hu2016baywatch,nelms2013execscent,oprea2018made,oprea2015detection}. However, these methods are usually used inside a single  organizational network and have limited capabilities of detecting attacks not seen at training time.

An important open question for thwarting global malware on the Internet is how to leverage defender collaboration and enable coordinated cyber defenses. To date, inter-organizational cooperation is used primarily for sharing threat intelligence in the form of Indicators of Compromise (IoC), such as IP addresses, domain names, and URL patterns used during an attack~\cite{johnson2016guide,wagner2016misp}. 
However, this approach has well-known limitations as it relies on detection of ongoing attacks and their associated IoCs, while attackers can change their infrastructure and behavior to make the detected IoCs obsolete~\cite{fereidooni2022fedCRI,mavroeidis2017cyber,tounsi2018survey}. 
This observation leads to the natural question: \emph{Are there other, more proactive and reliable approaches to global defense coordination
that could be effective against evolving, sophisticated cyber threats?}

In this work, we answer the above question affirmatively by presenting \thesystem, a federated machine learning framework for global HTTP-based threat detection. \thesystem\ enables collaboration among defenders to globally train neural network models for HTTP malware detection. 
The main insight in designing \thesystem\ is to facilitate \emph{knowledge transfer} among participants through a collectively trained model. A global machine learning model captures data diversity and thus becomes a powerful tool for uncovering a wider range of malware behavior characteristics.
This approach enables clients to detect malware seen for the first time in their networks, which a locally trained ML model will not identify.

In real-world deployments of ML threat detection systems, the vast majority of data is unlabeled, and  the ground truth of malicious samples is  limited. Additionally,  supervised defenses often fail to detect novel attacks not encountered in training, and it is difficult to get accurate labels of malicious activity on a network. To address these concerns, we propose the design of a new active learning component into the \thesystem\ federated learning framework with the goal of increasing the labeled set during the training phase and, eventually, enhancing the global model's detection capabilities. In our design, we use active learning to identify a small set of anomalous samples for investigation in each round of training, and augment the training set with maliciously labeled samples.

We evaluate \thesystem\ using a large dataset of HTTP logs collected at the border of two university networks. We also use three public datasets from different malware families  (Mirai, Gafgyt, and the data exfiltration malware dataset from \cite{bortolameotti2017decanter}), as well as attack recreation data generated on the university networks. We show that, across all the malware families we considered, the global model outperforms local models trained on a single client's data. 
The improvements obtained by global models are significant: 
For instance, the global model is able to detect a data exfiltration malware family with three times higher Precision Recall AUC (PR-AUC) than the local model.
Importantly, global defenses enable clients to detect new malware in their environments (i.e., malware that they have not seen in training), which is learned from the models shared by other clients participating in the federated protocol. We further demonstrate that when the active learning is enabled, \thesystem\ is able to detect \emph{completely new malware, for which labels are not available in training at any of the clients}.  In particular, the key intuition to enable new malware discovery is to include an anomaly detection module trained on each client network with the goal of detecting and labeling anomalies in the network. Often, new attack instances will result in anomalous traffic relative to the benign traffic. Once the anomalies are confirmed as attack instances, they are added to the next round of federated learning training, and the global model will learn to recognize these newly discovered attacks. 

One of the important threats against federated learning  is the adversarial manipulation by malicious or compromised clients to poison the model~\cite{bagdasaryan2020backdoor,sun2019can,fang2020local, shejwalkar2021manipulating, Wang21_Tails}.
We design a new defense technique specific to threat detection, called DTrust (Distributed Trust), with the goal of training in the presence of poisoning attacks by identifying and removing the malicious clients. The benign clients evaluate locally if the global model received from the server can be trusted, and notify the server if a large performance degradation is observed. The server investigation consists of inspection of individual model updates and identifying the clients that sent "bad" updates to remediate the attack.  DTrust relies on the insight that the most client organizations in a collaborative threat detection system act in good faith, and they are incentivized to actively participate in defense and validate the model during the training process to protect against poisoning attacks. We evaluate DTrust in three poisoning scenarios in which a number of compromised clients inject different attack patterns to be misclassified by the global model. In one scenario, the global model's PR-AUC degrades from $0.93$ to $0.11$ when no defense is deployed, but DTrust restores the PR-AUC to $0.93$ after identifying the malicious clients and removing them from training. 
Finally, we deploy \thesystem\ on  two university networks  using three attack recreation exercises performed at  intervals of several months.
We show that \thesystem\ detects the malicious communication carried out during the attack recreation with high precision and false positive rates lower than $3.3 \times 10^{-5}$ on both networks.
Moreover, the model trained during the first experiment  
still has strong performance on an evasive attack exercise performed four months later. 
In addition, \thesystem\ detected a set of 42 previously-unknown malicious URLs and 20 domains on the two networks, which were confirmed malicious by VirusTotal. 
To summarize, our contributions are:

\begin{itemize}[nosep]
    \item \textbf{Federated learning for global cyber defense:} 

    We design \thesystem, a scalable and privacy-preserving framework for federated training of global neural network models, which enables early-stage HTTP malware detection at participating organizations.

    \item \textbf{Active learning for limited ground truth:} We introduce a novel active learning component in our federated design, which selects samples for investigation and labeling using an anomaly detection module, and thus enables the discovery of completely new attacks.
    \item \textbf{Poisoning mitigation:} We design DTrust, a new poisoning detection and mitigation method specifically designed for federated learning in the collaborative threat detection domain. DTrust successfully detects poisoning clients using the feedback from participating clients to investigate and remove them from the training process.
    \item \textbf{Comprehensive evaluation:} We evaluate \thesystem\ using large-scale datasets from two university networks, public traces from three malware families, and three attack recreation exercises. We show that global models trained in \thesystem\ have high precision and recall, and low false positive rates,  improving significantly upon locally trained models. We further demonstrate the impact of poisoning attacks and effective mitigation with DTrust.

    \item \textbf{Deployment and detection of unknown malware:} We deploy \thesystem\ on  two university networks and find instances of previously-unknown malware (42 malicious URLs and 20 malicious domains). 
\end{itemize}

%% file: problem.tex
\section{Problem Statement and Threat Model}
\label{sec:problem}

In this section, we discuss the problem of HTTP malware detection, the limitations of existing solutions, and introduce our system requirements and threat model.

\vspace{2pt}
\BfPara{HTTP-based malware}
HTTP is one of the  most widely used  protocols  by adversaries to perpetrate malicious activities, such as data exfiltration~\cite{dataExfilt2020}, vulnerability exploitation~\cite{httpmirai2019}, and covert channel communication~\cite{httpstatusmalware2020}. Command-and-control (C2) servers often use HTTP to send commands to compromised systems and receive stolen data~\cite{sood2014empirical}.
Prior work proposed a variety of methods for detecting malicious network activity over HTTP, including URL-based detection~\cite{mamun2016detecting, mcgahagan2019comprehensive}, 
detection using web-proxy logs~\cite{bartos2016,hu2016baywatch,nelms2013execscent,oprea2018made,oprea2015detection}, and 
application fingerprinting~\cite{bortolameotti2017decanter, bortolameotti2020headprint, perdisci2010behavioral}. We focus on malicious activity detection using HTTP logs for several reasons. First, while malicious URL detection is shown to be successful in mitigating certain attacks (e.g., phishing), HTTP-based detection methods are able to address a variety of attack surfaces used by different malware families.  Second, the HTTP-based methods avoid deep inspection of the payload which could be obfuscated by attackers, and rely only on the HTTP headers for increased scalability and lower computation costs. Lastly, these approaches  utilize common HTTP logs collected by web proxies installed at the border of enterprise networks, which most enterprises' use as part of their perimeter control. 

Most of the existing machine learning detection techniques on HTTP logs attempt to detect malware activity within a single network by training local models on labeled data available in that network~\cite{hu2016baywatch,bartos2016, oprea2018made,oprea2015detection}. As attackers employ various techniques to evade detection, such as changing the C2 protocol, or changing the domain names and the IP addresses of the C2 infrastructures, local models will fail at detecting these. In this work, we address the problem of \emph{designing global detection of HTTP malware by coordination among multiple participating organizations}. We believe that collaboration among multiple defenders is critical in enabling a better and more resilient cyber defense strategy and will lead to more effective threat detection methods. Unfortunately, the state of the art in collaboration among multiple defenders is based on threat intelligence sharing platforms, which enable sharing of indicators of compromise (IoCs) after attacks are detected. IoCs have limited utility at detecting cyber threats, particularly as IoCs become stale with small changes to attackers' malicious infrastructures or communication protocols.  We are interested in more proactive, coordinated approaches among defenders that can resist evolving cyber attacks.

In designing our system, we identified several requirements  for real-world deployment in participating organizations: 
\begin{itemize}[nosep]
    \item \textbf{Globally coordinated detection}: We are interested in global methods that  expose attacks promptly, when they are still largely invisible to individual organizations.
    \item \textbf{Real-time processing}: We require real-time processing of HTTP logs to generate predictions on individual HTTP logs when the model is deployed. This approach enables early-stage attack detection and thus minimizes the damage inflicted by an attack. 
    \item \textbf{Scalability}: We would like to run the system on multiple networks with large volumes of logs. 
    \item \textbf{Log privacy}: The system should require minimal data sharing across the participating clients, in order to ensure the privacy of sensitive security logs.

    \item \textbf{High accuracy}: To be used in production, the system should have high accuracy, precision, and recall, as well as low false positive rates across different classes of HTTP-based malware attacks.
\end{itemize}

\BfPara{Challenges}
Detecting malicious communication over the HTTP protocol  is challenging for multiple reasons: 
First, the malicious traffic transmitted on the HTTP protocol is blending in with large volumes of legitimate traffic generated by users, making detection very difficult. Second, there is a high imbalance among the malicious and benign samples on a single network, making it challenging to train supervised learning methods with high accuracy and low false positive rates~\cite{SommerPaxson2010}.  Third, most existing systems for HTTP malware detection employ some form of aggregation of events from multiple logs. For instance, beaconing detection methods apply time series analysis on multiple communication events to the C2 server to detect periodic communication~\cite{hu2016baywatch}. It is challenging to detect malicious HTTP communication in real-time, as it requires generating a prediction on individual log events.

\vspace{2pt}
\BfPara{Threat Model}
We aim to design a globally coordinated system with multiple participating clients for  detection of  malicious activity over the HTTP protocol. The malicious communication could be part of multiple stages of a malware campaign, including malware delivery, the C2 communication with the malicious server, and data exfiltration activities.

We assume that the attacker compromises one or several victim machines within the participating  networks. Hosts in the network can be infected in a variety of ways, such as vulnerability exploits, social-engineering, and drive-by download attacks. The infection vector might not occur over HTTP and therefore, our system might not be able to detect the initial infection. However, our goal is to detect any subsequent malicious communication over HTTP, which is a common communication channel once attackers have established foothold in a network.

Our system analyzes HTTP logs collected at the border of the monitored networks.

The participating organizational networks  run local computation on the collected logs for training local models, and share those with a central server for aggregating a global model. 

We assume that the central aggregator correctly follows the protocol for aggregating the local updates into a global model. However, the clients participating in the protocol might be subject to data poisoning attacks (in which the logs they collect are under the control of an adversary), or model poisoning attacks (in which the adversary controls the updates sent to the central aggregator). Our goal is to train models that identify the malicious activity over HTTP, while providing resilience against poisoning.

%% file: method.tex
\section{System Overview and Methodology}

\input{method/overview}

\input{method/shortend_feat_emb}

\input{method/active_FL}

\input{method/poisoning}

%% file: method/overview.tex
In this section, we present an overview of \thesystem, followed by a detailed description of each system component.
\thesystem\ is a federated framework for cyber defense, where a set of participating clients collaboratively learn a global model $G$ from HTTP logs collected at their network border. The goal of the global model is to learn a classifier that generates predictions regarding whether individual HTTP logs are Malicious or Benign. We consider the \emph{cross-silo federated learning} setting, suitable
for a relatively small number of clients, in which all clients
participate in each round of training~\cite{kairouzAdvancesOpenProblems2021}.
Following the federated learning paradigm~\cite{mcmahan2017communication}, in a training iteration $t \in \left\{1,...,T\right\}$, each client $i \in \left\{1,...,n\right\}$ locally trains a local model $W_i^t$ based on the previous global model $G_{t-1}$, by performing stochastic gradient descent (SGD) updates on a subset of its local data, $D_i^t$. The clients send their local model updates to the server, which aggregates them to produce the updated global model $G_t$ and distribute it to the clients. The process continues iteratively until the global model converges. 

The training data (HTTP logs) $D_i$ maintained by each client $i$ locally is labeled as Malicious or Benign. 
We discuss in Section~\ref{sec:evaluation} how we generate ground truth for data labeling. 
Before training, the data undergoes a feature extraction phase where individual HTTP logs are processed to generate a set of 5862 features in three categories -- embedded, numerical and categorical (Section~\ref{sec:features}). 
We propose the use of embedded features from text-based HTTP fields (URL, domain and web referer) adapting word embedding representations from natural language processing (NLP). Word embeddings are generated with deep learning models such as Word2Vec~\cite{mikolov2013word2vec} and FastText~\cite{bojanowski2016enriching} that are trained to capture contextual and semantic similarities in the text, such that similar words will also be closer in the embedded vector space. 

\thesystem\ uses federated learning techniques for two tasks: (1) generating embedded feature representations, and (2) training a neural network classifier for detection of malicious HTTP traffic.
For the first task, we introduce a federated method in which each client updates sequentially a shared embedding model using their own data corpus. For the second task, clients use their local HTTP feature vectors  to collectively train a global model that is able to detect various attack patterns. 
\thesystem\ introduces  a novel  Active Learning component that uses a local anomaly detection module to augment the ground truth of malicious activities through sample selection (Section~\ref{label:activeFL}). In addition, \thesystem\ employs a novel distributed defense mechanism against poisoning attacks, in which the clients themselves help identify malicious attempts to corrupt the global model (Section~\ref{label:pois}). 
We detail each component of \thesystem\ in the following sections.

%% file: method/shortend_feat_emb.tex
\begin{figure*}[!t]
\centering
\includegraphics[width=0.95\linewidth]{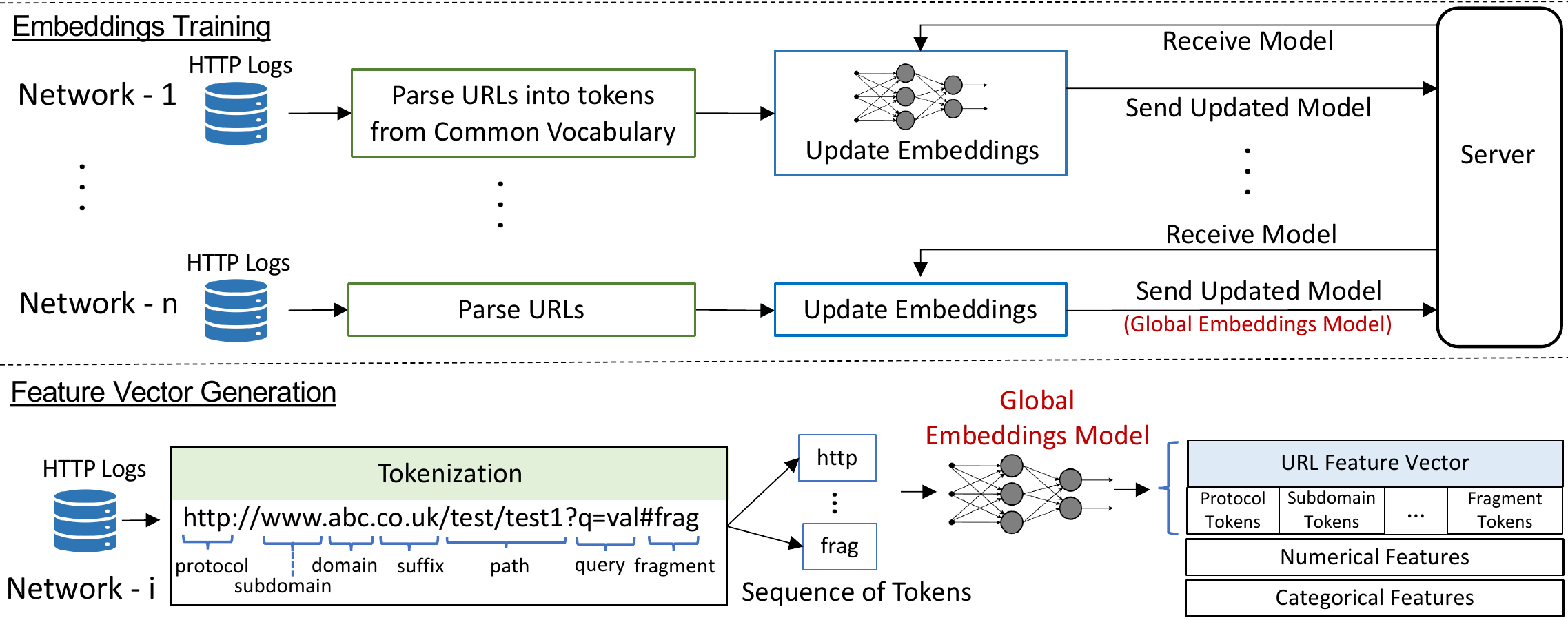}
\caption{Federated embedding model training for URL representation. We generate embedded features for URL, Domain and Referer. In addition to embedded features, we also include numerical features and categorical features.}
\label{fig:embeddings}
\vspace{-10pt}
\end{figure*}

\subsection{Feature Representation}
\label{sec:features}
Our system is unique in its requirement for real-time processing of HTTP log events. Thus, timing features and aggregated features over multiple HTTP logs or flows are not applicable in our setting. 
We leverage all the available fields in HTTP logs, except the source IP address (since our detection methods are not specific to the host machine), and source port (which does not carry information related to our detection task). We include three feature categories: \emph{embedding features} for domain, URL, and web referer representation, \emph{categorical features} for external IP subnet, port, user agent string, method, status code, and content type, and \emph{numerical features} for request and response size, transaction depth, and browser version. We summarize all features in Table~\ref{tab:features} from Appendix~\ref{apx:features}.

URLs are one of the HTTP fields particularly susceptible to adversarial manipulation; they are often used by malware to communicate information with C2 servers, through various parts of the URLs, including sub-domain, parameters, query string, file name, and fragment. Hence, it is important to have an effective, semantics-aware feature representation of URLs.
Different from prior work on URL feature representation~\cite{blum2010, ma2009beyond, mamun2016detecting,khramtsova2020federated,saxe2017expose,le2018urlnet}, we propose an embedding model which preserves the semantic structure of the URL, handles new tokens at deployment time, and can be trained in a federated fashion. 

In Figure~\ref{fig:embeddings}, we present our design for creating the embedding representation in a federated unsupervised manner. Initially, we parse the URLs into tokens representing different categories (e.g., domain, path, query string) to preserve the structure of an URL. Each token is considered a word, and each URL is viewed as a sentence of words. 
Our contribution is the design of a more scalable and privacy-preserving approach for training the embeddings model among multiple participants via distributed sequential client updates. In this approach, each client has access to its own data and updates the global embeddings model locally, using its entire corpus. The client sends the updated global model back to the server, who acts as a trusted central coordinator. The clients apply their updates sequentially, in a round robin fashion, over multiple iterations, as illustrated in the top part of Figure~\ref{fig:embeddings}. \thesystem\ uses federated FastText as the preferred embedding method for URL representation. Compared to Word2Vec, FastText has the advantage that the dictionary is a set of n-grams, hence it is well known to all participants in the protocol and does not need to be shared. Additionally, FastText  embeddings can handle new tokens encountered at deployment time.  Other considerations on our design choices for feature representation are described in Appendix~\ref{apx:features}.

%% file: method/active_FL.tex
\subsection{Active Federated Learning}
 \label{label:activeFL}
In this work, we employ the commonly used Federated Averaging (FedAvg)~\cite{mcmahan2017communication} training method, where the client models are weighted by their dataset size and averaged to update the global model as follows: $G_t=\sum_{i=1}^{n}\frac{||D_i^t||}{||D^t||} \times W_i^t$, where $||D^t||=\sum_{i=1}^{n}||D_i^t||$.
The local training on each client is carried out using a Feed-Forward Neural Network (FFNN) model based on our feature representation.

Training supervised models such as FFNN requires labeled data, a known challenge in cyber security~\cite{SommerPaxson2010} where the vast majority of data available in real-world deployments is unlabeled. Furthermore, existing defenses that might work well on previously seen attacks often fail at detecting new emerging threats, making it particularly challenging to get accurate labels of malicious activity on a network. To address these concerns, we integrate an active learning component into \thesystem\  with the goal of augmenting the ground truth of malicious activities incrementally, at each iteration of the training phase. Thus, the global model's detection capabilities are improved in a streaming fashion, similar to online learning where true class labels of new incoming samples are usually unknown~\cite{lughofer2012single}. This approach fits particularly well with federated learning, which is designed for continuous training over time, and thus can account for malware evolution over time. To the best of our knowledge, \thesystem\ is the first threat detection system using active learning in a federated method of training with limited labeled data.  

\begin{figure}[thbp]
	\centering
	\includegraphics[width=1\linewidth]{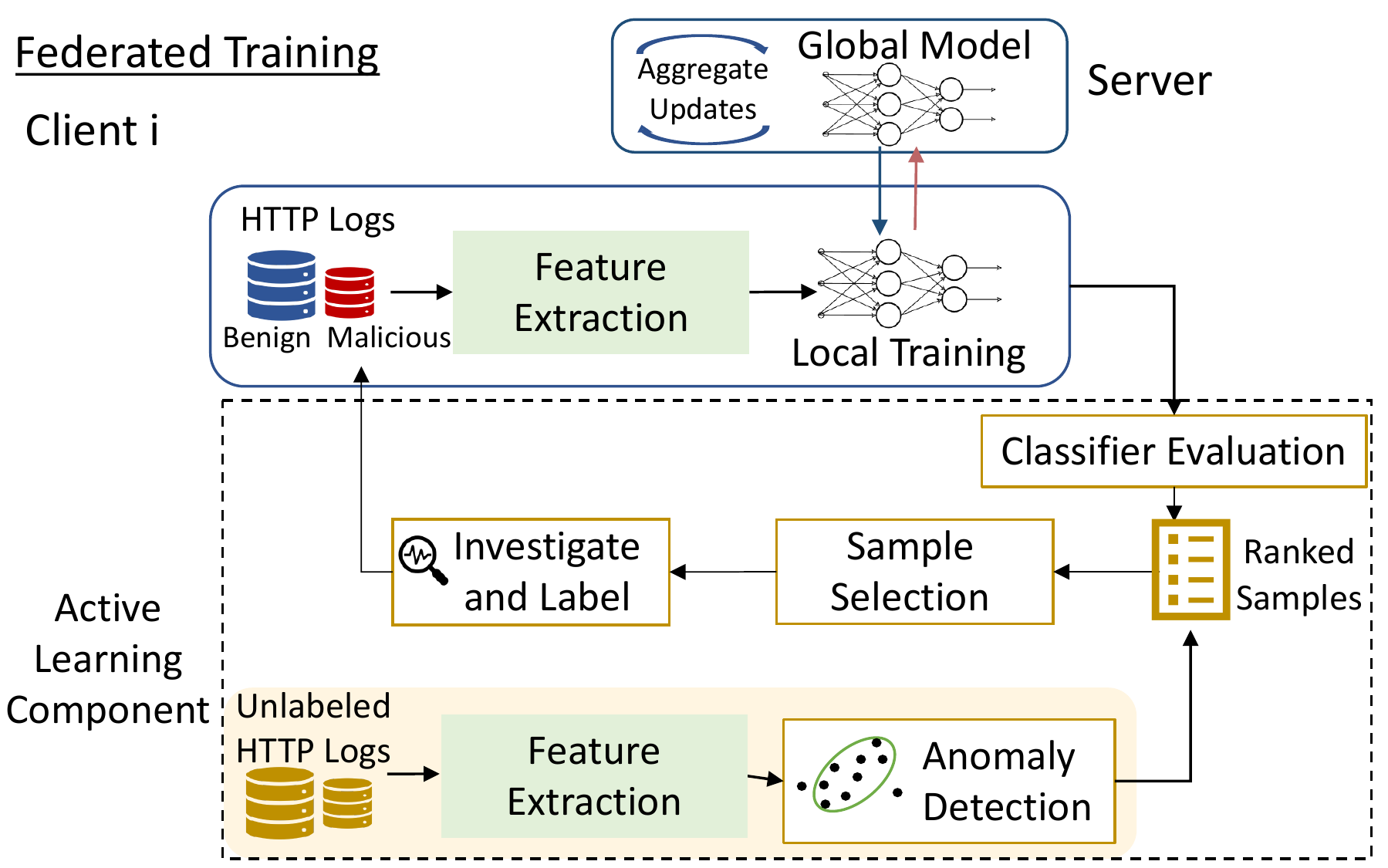}
	\caption{The Active Federated Learning framework. At time step $t$, the global model $G_t$ is used to rank the unlabeled data of client $i$. In addition, an anomaly detector is used to rank the remaining unlabeled data samples by their anomaly score. The top ranked samples are investigated and labeled by a human expert and used in the next iterations.}
	\label{figure:active_fed}
	\vspace{-10pt}
\end{figure}

Several strategies for selecting instances to be labeled in active learning have been previously explored, such as random sampling, uncertainty sampling~\cite{lewis1994}, expected error reduction~\cite{Roy2001TowardOA}, and variance reduction~\cite{CohnEtAl:96}.
In our case, we are interested in expanding the malicious ground truth, therefore we combine both uncertainty sampling and anomaly detection in a hybrid approach for sampling the most suspicious HTTP log events to be investigated, as proposed by prior work~\cite{stokes2008aladin}.  
The Active Federated Learning mechanism is illustrated in Figure~\ref{figure:active_fed}. 
At time step $t$, the global model $G_t$ is used to rank the unlabeled data of client $i$. In addition, an anomaly detector is used to rank the remaining unlabeled data samples by their anomaly score. The top ranked samples are investigated and labeled by a security analyst and used in next iterations.
We consider a small budget $b$ of samples that are investigated and labeled by a human expert at one of the participating organizations $i$, and time step $t$. 
The first strategy consists of using the global model $G_t$, represented by a Feed-Forward Neural Network classifier, to evaluate and rank the unlabeled dataset; the most suspicious $b/2$ samples are then investigated and labeled by a security analyst. These are likely similar to previously seen malicious samples. The second strategy consists of using a local anomaly detection module to rank the remaining unlabeled data according to an anomaly scoring method; the most anomalous $b/2$ samples are then investigated and labeled by a security analyst. The second strategy identifies anomalous logs that could augment the ground truth with new attack samples not seen in training before. The combination  of the two strategies is critical for \thesystem's ability to detect attacks with very limited ground truth. 
The samples labeled malicious are added back to the training dataset of client $i$ and used in the next iterations of federated learning. We focus on labeling malicious samples only, as some adversarial actions might look benign during stealthy attacks. We assume the human analysts make the correct labeling decision, and note that learning from noisy labels is an extensively studied ML topic on its own, not specific to our system~\cite{Frenay2014,Johnson2022ASO}.

We demonstrate in our evaluation section that active federated learning can be very powerful in some scenarios, even when starting the FL training process with zero labels.
Our anomaly detection module uses the well-known Isolation Forest algorithm~\cite{liu2008isolation}, however other anomaly detection methods such as local outlier factor (LOF)~\cite{breunig2000lof}, one-class support vector machines (SVM)~\cite{Scholkopf2001}, and clustering\cite{Campello2015} can also be employed.
To increase resilience, we train an ensemble of Isolation Forest models on multiple time windows, where each time window corresponds to one FL iteration (time step). Specifically, given a client $i$, the anomaly detector is trained on unlabeled data from the previous $k$ time windows (i.e., the datasets $D_i^{t-k}, ... D_i^{t-1}$), while anomaly detection is carried out on the current time window $t$ (i.e., the dataset $D_i^{t}$). 
For instance, a good trade-off between speed and performance was reached in our experiments at $k = 3$.

%% file: method/poisoning.tex
\subsection{Resiliency to Poisoning Attacks}
\label{label:pois}

Due to its distributed nature, federated learning is vulnerable to adversarial manipulation by malicious or compromised clients. Poisoning attacks that corrupt the training data~\cite{tolpegin2020poisoning} or the client model updates~\cite{bagdasaryan2020backdoor,sun2019can,fang2020local, shejwalkar2021manipulating, Wang21_Tails} have been an important attack vector studied in recent years.
 Existing defenses  can be classified as: (i)  defenses that limit the contribution of each client to the global model~\cite{sun2019can} or perform anomaly detection to remove specific client updates~\cite{blanchardMachineLearningAdversaries2017,mhamdiHiddenVulnerabilityDistributed2018,yinByzantineRobustDistributedLearning2018a}; (ii) defenses against  backdoor poisoning that detect and mitigate the presence of a trigger backdoor~\cite{rieger2022deepsight,nguyen2022flame}; (iii) defenses that leverage a trusted dataset at the server to filter malicious updates~\cite{DBLP:conf/ndss/CaoF0G21}. We are interested in developing a general defense against both data and model poisoning attacks, that leverages the specifics of our threat detection setting.

\vspace{2pt}
\noindent\textbf{DTrust algorithm:}
We propose a novel defense algorithm called DTrust (Distributed Trust), with the goal of training a resilient federated model in the presence of poisoning attacks.
In essence, DTrust is a distributed algorithm where the benign clients locally evaluate if the global model received from the server can be trusted, and notify the server if a large performance degradation is observed.
We introduce a novel distributed scheme of bootstrapping trust, in which the clients take an active role in detecting poisoning attacks. Each client $i$ interested in participating in \thesystem's DTrust defense maintains a small dataset $D_i^{Trust}$ called the \emph{trust dataset}, which they share with the server in case performance degradation is noticed.  Our insight is that multiple clients having individual trust datasets cover a wider data range with more opportunity for attack detection; intuitively, one attack that is undetected by one client might stand out on another client's trust dataset. 
DTrust is particularly suited for our threat model, where the clients are organizations (rather than IoT devices or individual machines), hence they have enough resources to actively participate in defense. 

DTrust fully exploits the distributed nature of FL, in contrast to  centralized trust-based schemes that rely on the server to detect adversarial attempts~\cite{fang2020local,DBLP:conf/ndss/CaoF0G21}. 
In \thesystem\, clients can maintain and share their trust datasets with the server, and the server   
can leverage publicly available tools (e.g., VirusTotal, third-party threat intelligence platforms, etc.) to ensure that the trust datasets are clean. For instance, the trust dataset could contain command-and-control traffic to a domain, where the domain has a high score on VirusTotal. Such tools may not be readily available in other application domains (e.g., image classification, human activity recognition, etc.)  where trust-based defenses have been explored~\cite{DBLP:conf/ndss/CaoF0G21}, making this approach particularly feasible for our cyber security setting.

A single iteration of DTrust at time step $t$ for client $i$ and the server can be summarized as follows.
When a client receives the global model $G_t$ from the server, it evaluates the model on its trust dataset. In our implementation, we use the cross-entropy loss metric $L$, although other performance metrics could also be used (e.g., error rate, PR-AUC score, F1-score). The \emph{loss impact} at client $i$ is defined as $\frac{L - L_{min}}{L_{min}}$, where $L, L_{min}$ are the current loss and the minimum loss across previous iterations, respectively. 
If the loss impact exceeds a threshold $T$, then the client notifies the server and sends the iteration number $t_{best}$ of the previous best-performing global model and its trust dataset (along with its model updates) to the server for investigation.

The server maintains a history database $H$, containing: (1) the previous global models $G_1, ..., G_{t-1}$, and (2) the most recent model updates $W_i^{t-1}$ received from each client $i$ at time step $t-1$.
When the server is notified of a large performance degradation by one of the clients, it uses that client's trust dataset to investigate which update from the previous round of training $t-1$ has caused the performance drop, as follows:
\begin{enumerate}[nosep]
        \item Confirm that the trust dataset sent by the client is legitimate, i.e., contains malicious samples that can be verified using threat intelligence tools. 
        \item Validate that the global model at $t_{best}$ performs well on the trust dataset, whereas the current model is significantly worse. 
        \item Leave one client $m$ out and aggregate a global model $G_{t-1}'$ without $m$'s updates. 
        \item Evaluate the global model $G_{t-1}'$ on the trust dataset received. 
        \item Compute the loss impact caused by client $m$ as $\frac{L_{t-1}-L_{t-1}'}{L_{t-1}'}$, where $L_{t-1}, L_{t-1}'$ represent the loss of the global model with and without client $m$, respectively.  
        \item If the loss impact exceeds a threshold, conclude that client $m$ may be poisoned; remove it from the set of clients, and revert the model to exclude its updates.
        \item Communicate the observation with client $m$, to investigate the incident.
\end{enumerate}

DTrust can be extended to account for performance degradation across multiple rounds of training. Instead of evaluating the loss impact of client $m$'s last update ($t-1$), the server would evaluate the loss impact of $m$'s last $k$ updates, which have been stored in history. This approach increases DTrust's resilience to stealthy poisoning attacks that happen across multiple training time steps.

%% file: evaluation.tex
\section{Evaluation}
\label{sec:evaluation}

\input{evaluation/settings}

\input{appendix/eval_features}
\input{evaluation/eval_federated}

\input{evaluation/eval_active}

\input{evaluation/eval_poisoning}

\input{evaluation/eval_deployment}

%% file: evaluation/settings.tex
\BfPara{Datasets}
We evaluate our system on a number of data sources. Each client organization participating in \thesystem\ maintains a labeled set of malicious and benign samples. 
We use the two university networks' traffic as benign data, and use various malicious data sources for different experiments as explained below.

\emph{(A) University Network Dataset:}
We obtained access to HTTP logs from two large university networks. The networks contain an average of 20 and 9 million HTTP log events per day, respectively, and were used actively during the period of data collection. 
Sensitive fields in the logs (such as internal IP addresses and URL parameter values) are anonymized in a consistent manner to protect user  personal information. We performed all our analysis on servers within
the university network. The IRB office at one of the universities reviewed our data
collection process and determined that this research does not qualify as Human Subject Research. However, our team members participated in IRB training and used best practices to handle the network data. 
We filter this dataset based on domain popularity using Tranco~\cite{pochat2018tranco} to exclude top 10,000  domains. We further remove external domains contacted more than 10,000 times. Both of these sets of domains are unlikely to serve malware, and similar methodology was used in previous work for data reduction~\cite{hu2016baywatch,oprea2018made}.  After filtering the traffic, we have between 1,900 and 31,000 logs per 30-minute time interval. 

\emph{(B) Network IDS logs (NIDS)}
In one of the university networks, a commercial off-the-shelf network IDS is available, and alerts are collected in real-time for potentially malicious events. We have access to the historical samples, and we collect malicious domains and IP addresses from the malware-callback alerts. We process 28,985 alerts, containing 9,124 unique requests and 687 unique hosts (domain and IP addresses). We use the collected indicators to match and label connections in each network as malicious during the training period.

\emph{(C) IoT Malware Dataset (Mirai, Gafgyt)}
We collect a dataset of IoT malware (Mirai and Gafgyt) from CyberIoCs, VirusTotal, VirusShare, and malicious samples shared by Alrawi~\etal~\cite{AlrawiLVC21}. We build a dynamic analysis framework to execute the malware in a sandboxed environment to collect the HTTP logs for each malware family. In total, we use 300 malware samples from each family, containing 20,627 and 13,889 HTTP events. 

\emph{(D) Data Exfiltration Malware Dataset (DEM)}
Bortolameotti et al.~\cite{bortolameotti2017decanter}  collected HTTP logs for data exfiltration malware samples in a sandbox and released the dataset. This dataset contains multiple malware families: Shakti, FareIT, CosmicDuke, Ursnif, Pony, Spyware, and SpyEye, and has more variability. In total, 79 malware samples containing 8,843 HTTP events are used in the experiments. In several experiments, we merge the traffic of IoT  and DEM malware  at training and testing time to create controlled experiments with known ground truth of malicious activities. 

\emph{(E) Attack Recreation Dataset (ARD)}
To conduct a realistic experiment for evaluation, we set up an attack recreation infrastructure  and run controlled attacks on the two networks. We vary the C2 infrastructure, including the C2 domains and IP addresses, to emulate adversarial evasion capabilities and test the performance of \thesystem. During attack recreation, we record malicious labels on individual HTTP logs. Therefore, this setup offers fine-grained ground truth, which enables us to evaluate the performance of \thesystem\ in a real environment. The details of the exercises are given in Section~\ref{sec:deployment}.

\BfPara{Implementation}
We implemented the system in Python 3.8, and used {\it Keras~\cite{chollet2015keras}} and  {\it TensorFlow Federated~\cite{tensorflowfederated}} for federated training. We use the  {\it Gensim} framework~\cite{vrehuuvrek2019gensim} for training embedding models.
There are several hyper-parameters which need to be selected before training. Federated learning updates happen within rounds, and we set the duration of a round to $t=30$ minutes. Parameter $t$ is tuned to balance the trade-off between communication cost and convergence speed: increasing the training time of each round results in smaller communication cost overall, but slows down the convergence. The neural network hyper-parameters are selected after a grid-search: we use a single hidden layer of size $128$, a dropout layer with rate $0.1$, and learning rate of $0.01$.

\BfPara{Evaluation Metrics}
We evaluate our system using Precision and Recall metrics, as our dataset is also imbalanced. 
We measure the performance with Precision-Recall Area Under Curve (PR-AUC), a single metric that captures the performance across all thresholds. We also use False Positive Rate (FPR) as an important metric for the models to be deployed in a real-world setting.


%% file: appendix/eval_features.tex
\subsection{Feature Set Comparison}
\label{apx:exp_features}

\begin{table}
	\centering
\begin{tabular}{c|cccc}
\begin{tabular}[c]{@{}c@{}}\textbf{URL}\\\textbf{ Representation}\end{tabular} & \textbf{Precision} & \textbf{Recall} & \textbf{F1} & \textbf{PR-AUC}  \\ 
\hline
Word2vec (federated)                                                           & 0.96               & 0.78            & 0.86        & 0.88             \\
FastText (federated)                                                           & 0.96               & 0.79            & 0.87        & 0.88             \\
Word2vec (centralized)                                                         & 0.96               & 0.78            & 0.87        & 0.88             \\
FastText (centralized)                                                         & 0.95               & 0.78            & 0.86        & 0.88             \\
Lexical Features~                                                              & 0.94               & 0.69            & 0.80        & 0.83            
\end{tabular}
\caption{Feature representation comparison for detecting malicious URLs. Our federated approach using Word2Vec and FastText is similar to centralized embeddings, and they outperform the baseline lexical features.} 
\label{tab:emb_comparison}
\vspace{-10pt}
\end{table}

\BfPara{URL Representation}
We evaluate our embedding-based approach for malicious URL detection in both centralized and federated settings. We also evaluate a previous approach using lexical features~\cite{mamun2016detecting, khramtsova2020federated}, in order to compare with the embedding representation.  

\noindent \textbf{Embedding Model Training.} We train both Word2Vec and FastText models in centralized and federated settings. We use the domain and URL fields of HTTP logs to evaluate different representations. We train on one day containing 537,448 background samples, together with 18,853 malicious samples from the three malware families (Mirai, Gafgyt, and DEM). We test on the next day with 89,139 background samples, and 4,267 malicious samples. The training corpus consists of 7,060,904 URL instances, and a vocabulary of 995,059 unique tokens. The size of the embedding vectors is 32, and we use Continuous-Bag-Of-Words, with a context window of size 5. We selected these hyper-parameters after experimenting with several choices. 
We compare the following embedding methods: (1) Centralized Embeddings: The training is performed at the server using the data shared by the clients, and (2) Federated Embeddings: Each university network is a client, which gets access to its own data and performs the federated embedding protocol with the server.

\BfPara{Lexical Features} We use the code  by Khramtsova et al.~\cite{khramtsova2020federated} to generate  72 lexical features (e.g., length and entropy of each URL category, number of letters in each category, etc.). 

Table~\ref{tab:emb_comparison} shows the comparison of different feature representations in terms of precision, recall, F1 scores, and PR-AUC. We make three observations: First, we note that FastText provides similar performance to Word2Vec, and the training time on our  corpus is comparable. Since FastText provides better generalization by  handling out-of-dictionary words,  we choose FastText as the default URL embedding method in our framework. Second, we note that the federated embedding models provide similar performance to the centralized models, and we used this approach due to its increased privacy assurances. 
Third, we see that the lexical feature representation provides  worse performance on our data, with 10\% lower recall and similar precision compared to the federated FastText method. This result implies that the embedding method generalizes better at detecting a larger number of malware samples, which can evade the lexical features. The cost of training the federated embedding is 75 minutes in our setup.
 
\begin{table}
	\centering
\begin{tabular}{c|cc}
\textbf{Feature Groups} & \textbf{PR-AUC} & \textbf{Training Time}  \\ 
\hline
All\_Features           & 0.95            & 36m 55s                 \\
Domain\_URL             & 0.88            & 13m 23s                 \\
UA\_Referer             & 0.53            & 24m 6s                  \\
External\_Host          & 0.63            & 8m 4s                   \\
HTTP\_Metadata          & 0.52            & 8m 1s                  
\end{tabular}
\caption{Feature group comparison for detecting malicious HTTP events. We use four feature groups and compare to all our extracted features using  PR-AUC. All combined features achieve highest PR-AUC at detecting malware.}
\label{tab:feat_comparison}
\vspace{-10pt}
\end{table}

\BfPara{Feature Groups}
We investigate here the need for additional features extracted from HTTP logs to augment the URL embeddings. URLs are not always explicitly used in malicious communication, and other HTTP fields, such as IP address, referer and user agent string could also be attack indicators. We  compare several feature groups, as defined below:
 
\begin{itemize}[nosep]
\item {\bf Domain\_URL}: Embeddings of domain and URL.
\item {\bf UA\_Referer}: User agent categorical features and referer with embedding representation.
\item {\bf External\_Host}: External IP and port represented as categorical features.
\item {\bf HTTP\_Metadata}: Method, status, content type as categorical features, and trans depth, request length, response length as numerical features.
\item {\bf All\_Features}: Full feature vector from Table~\ref{tab:features}.
\end{itemize}

For this experiment, we used the same dataset as for URL detection, but with full HTTP logs.  
Table~\ref{tab:feat_comparison} shows the  PR-AUC for these feature groups. The results demonstrate that the domain and URL features have the highest performance in the 4 considered feature groups (PR-AUC of 0.88), but using all HTTP features improves PR-AUC to 0.95. There is an increase in the training time using all features, due to the larger dimension of the neural network model.

%% file: evaluation/eval_federated.tex
\subsection{Federated Models vs Local Models}
\label{sec:eval_fl}

\begin{table*}
\centering
\begin{tabular}{cc|cccc|cccc}
\multicolumn{2}{c|}{\textbf{Training Malware}} & \multicolumn{4}{c|}{\textbf{Deployed Network: UNI-1}}                                                                                                                                                                                                                          & \multicolumn{4}{c}{\textbf{Deployed Network: UNI-2}}                                                                                                                                                                                                                           \\ 
\hline
\textbf{UNI-1} & \textbf{UNI-2}                                                                    & \textbf{Test } & \begin{tabular}[c]{@{}c@{}}\textbf{Local }\\\textbf{\textbf{PR-AUC}}\end{tabular} & \begin{tabular}[c]{@{}c@{}}\textbf{Global }\\\textbf{\textbf{PR-AUC}}\end{tabular} & \begin{tabular}[c]{@{}c@{}}\textbf{Global FPR}\\\textbf{~at Recall 0.9}\end{tabular} & \textbf{Test} & \begin{tabular}[c]{@{}c@{}}\textbf{Local }\\\textbf{\textbf{PR-AUC}}\end{tabular} & \begin{tabular}[c]{@{}c@{}}\textbf{Global }\\\textbf{\textbf{PR-AUC}}\end{tabular} & \begin{tabular}[c]{@{}c@{}}\textbf{Global FPR}\\\textbf{~at Recall 0.9}\end{tabular}  \\ 
\hline
Mirai          & Gafgyt                                                                            & Gafgyt         & 0.8751                                                                            & 0.934                                                                              & 0.002                                                                                & Mirai         & 0.8806                                                                            & 0.9591                                                                             & 0.002                                                                                 \\
Mirai          & DEM                                                                               & DEM            & 0.2647                                                                            & 0.7522                                                                             & 0.01                                                                                 & Mirai         & 0.5483                                                                            & 0.8774                                                                             & 0.002                                                                                 \\
Gafgyt         & Mirai                                                                             & Mirai          & 0.8952                                                                            & 0.9308                                                                             & 0.003                                                                                & Gafgyt        & 0.9055                                                                            & 0.9568                                                                             & 0.001                                                                                 \\
Gafgyt         & DEM                                                                               & DEM            & 0.2429                                                                            & 0.7548                                                                             & 0.01                                                                                 & Gafgyt        & 0.4464                                                                            & 0.8016                                                                             & 0.002                                                                                 \\
DEM            & Mirai                                                                             & Mirai          & 0.7231                                                                            & 0.8508                                                                             & 0.003                                                                                & DEM           & 0.5483                                                                            & 0.7958                                                                             & 0.009                                                                                 \\
DEM            & Gafgyt                                                                            & Gafgyt         & 0.6013                                                                            & 0.7983                                                                             & 0.005                                                                                & DEM           & 0.4736                                                                            & 0.7939                                                                             & 0.009                                                                                
\end{tabular}

\caption{Federated model comparison with the locally trained models for detecting  malware families in multiple scenarios  with different training sets. Each row shows the training malware family in each network, and the PR-AUC for detecting the malware that is not locally seen during training. The results show the PR-AUC improvements of the global models compared to locally trained models and the low FPR of the global models.}
\label{tab:federated_vs_local}
\vspace{-10pt}
\end{table*}

 In this section, we investigate how a network can benefit from federated training when a new malware has been previously detected in one of the collaborating networks, but is seen locally for the first time.
We design experiments to use different malware families in the two networks during training, and show how the global model helps detection of a malware family seen for the first time in one of the networks. In particular, we generate a set of controlled experiments with the two university networks as clients. We merge malware traces from the three families (Mirai, Gafgyt, and DEM) in both networks to determine the performance of the local and global models under multiple scenarios. We split each of the malware family into 80\% samples for training and 20\% for testing, and distribute them to the networks across rounds.
 
 Table~\ref{tab:federated_vs_local} shows the PR-AUC  of the global models trained with federated learning and the local model in various settings. The results show the improvements achieved by the global model, which successfully transfers the detection from one network to other participating clients. For example, when Mirai and DEM malware families are used for training at UNI-1 and UNI-2, respectively, the federated  model achieves a PR-AUC of 0.75 at detecting DEM in UNI-1, a significant improvement over the local model's 0.26 PR-AUC. Similarly, at UNI-2, the global model achieves a PR-AUC of 0.87, compared to the local model's PR-AUC of 0.54 at detecting Mirai. In other cases, malware families share some common characteristics, which aids the local detection. In the case of Mirai and Gafgyt, a local model that trains on Mirai and later attempts to detect Gafgyt (or the other way around) performs relatively well (at 0.87-0.9 PR-AUC). Nonetheless, the global model outperforms it consistently, achieving a PR-AUC of about 0.93-0.95. False positive rates are consistently low for the global models, between 0.002 and 0.01, at a recall of 0.9.

\begin{figure}[t]
	\centering
		\includegraphics[width=0.85\linewidth]{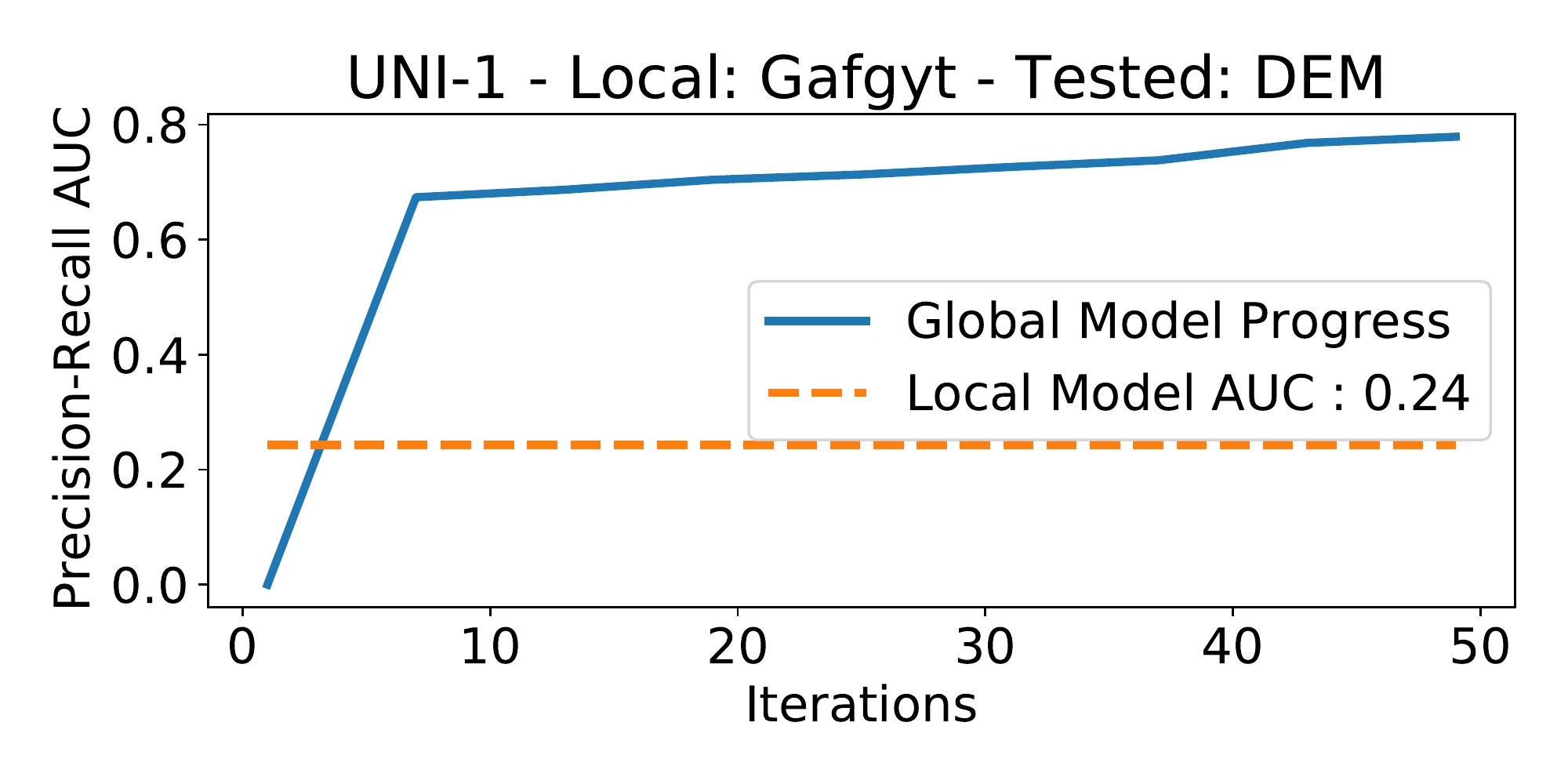}
 	\vspace{-8pt}

		\includegraphics[width=0.85\linewidth]{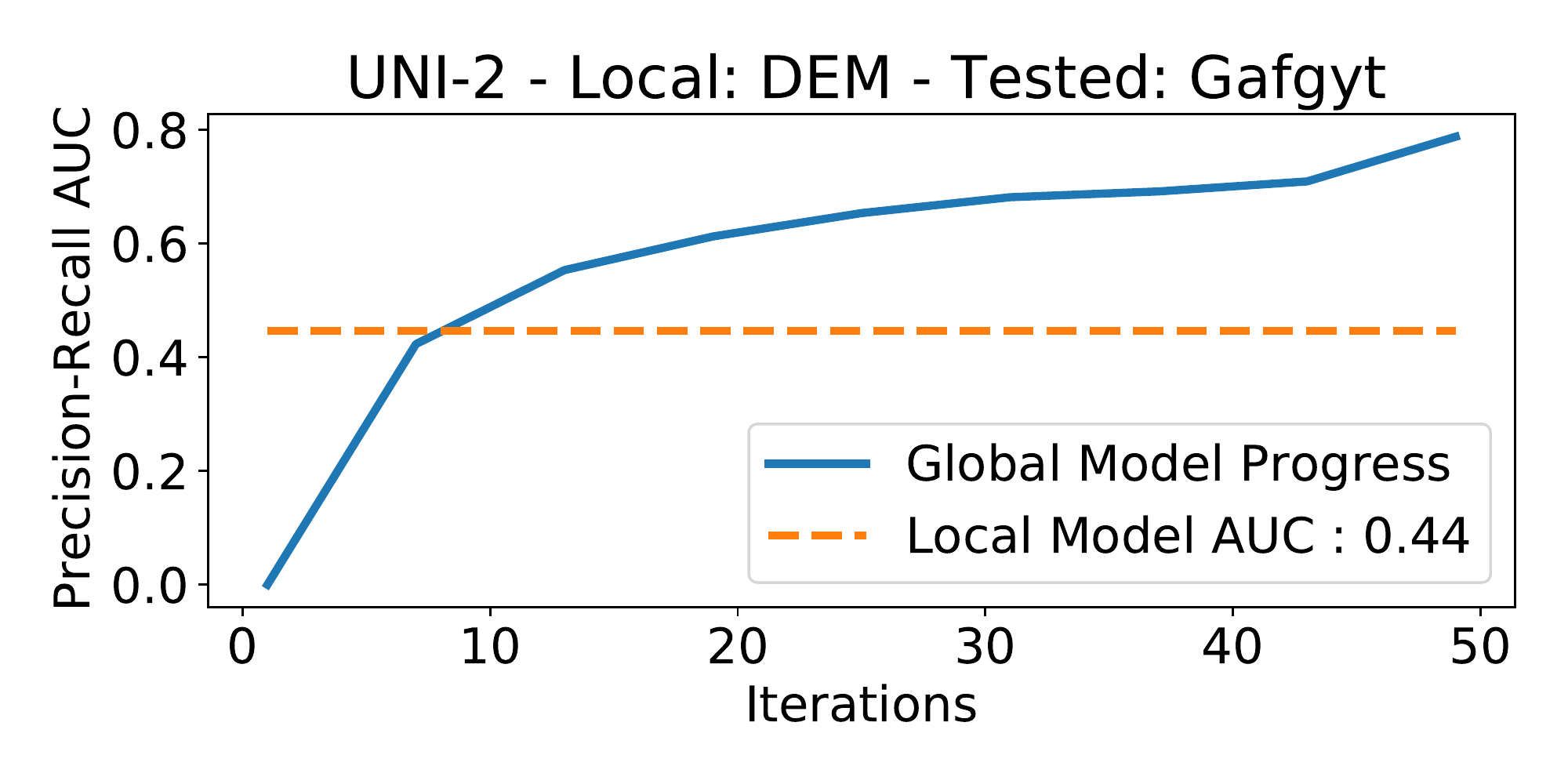}
   	\vspace{-8pt}

	\caption{Federated model progress for detecting malware families over multiple iterations 
 evaluated on the two networks. The two sites are seeing a new malware for the first time (DEM at UNI-1 and Gafgyt at UNI-2) and attempt to detect it with their own local model and with the federated model. Each iteration processes 30 minutes of data, and the global model exceeds the local model performance in less than 10 iterations.} 

	\label{fig:federated_vs_local_pr} 

\end{figure}

 Figure~\ref{fig:federated_vs_local_pr} shows the progress during federated training for two of the experiments. As more training data is consumed, the PR-AUC increases and the global model outperforms the local model in less than 10 iterations (5 hours), illustrating a clear performance gain with the federated  model.
 
\subsection{Scaling to Multiple Clients}
\label{sec:eval_scale}

\begin{table}
	\centering
\begin{tabular}{c|ccc|ccc}
                                                                          & \multicolumn{3}{c|}{\textbf{Network: UNI-1}}    & \multicolumn{3}{c}{\textbf{Network: UNI-2}}      \\ 
\cline{2-7}
\begin{tabular}[c]{@{}c@{}}\textbf{Global }\\\textbf{Model }\end{tabular} & \textbf{Mirai} & \textbf{Gafgyt} & \textbf{DEM} & \textbf{Mirai} & \textbf{Gafgyt} & \textbf{DEM}  \\ 
\hline
4 Clients                                                                 & 0.96           & 0.90            & 0.71         & 0.94           & 0.88            & 0.59          \\
10 Clients                                                                & 0.96           & 0.93            & 0.72         & 0.95           & 0.91            & 0.64          \\
16 Clients                                                                & 0.96           & 0.94            & 0.74         & 0.95           & 0.92            & 0.69          \\
20 Clients                                                                & 0.95           & 0.90            & 0.70         & 0.95           & 0.93            & 0.64          \\
24 Clients                                                                & 0.96           & 0.95            & 0.74         & 0.95           & 0.94            & 0.66          \\
30 Clients                                                                & 0.95           & 0.89            & 0.81         & 0.94           & 0.91            & 0.70         
\end{tabular}
	\caption{Federated model PR-AUC as more clients participate during training.  Each row shows the total number of clients, and the  PR-AUC for detecting different malware families in the two networks. As the number of clients increases and more data is used during training, overall detection performance generally improves.}
	\label{tab:federated_multi_client}
\end{table}

\thesystem\ is designed for multiple clients, but in cyber settings it is challenging to obtain data from multiple organizations.
We have access to  two large university network logs, and we use subnet-level information in each network to create multiple clients. For this experiment, we generate all the subnets at each university network, and create 15 clients in each network, each using data from several subnets.  We then consider a subset of these clients, varied between 2 and 15 at each network, to test the impact of the number of clients participating in the FL training protocol. We split the malware samples from the three families equally across the 30 clients in both networks. 

Table~\ref{tab:federated_multi_client} shows the global model PR-AUC at detecting the three malware families on the two university networks. We observe that for Mirai, four clients is enough to obtain good performance (PR-AUC above 0.94), whereas for DEM and Gafgyt families the performance improves as more clients participate in the protocol and more malicious data is available. The improvements are particularly large for DEM, which consists of multiple malware families and has more variability. The experiment shows that the system can scale to multiple clients and its detection performance improves with more clients participating in the FL protocol. Small variations in performance are due to differences in data distributions of clients joining the training process.

%% file: evaluation/eval_active.tex
\subsection{Active Federated Learning}
\label{sec:eval_active}

\begin{table}[t]
\centering
\begin{tabular}{c|ccc|ccc}
	& \multicolumn{3}{c|}{\textbf{Network: UNI-1}}    & \multicolumn{3}{c}{\textbf{Network: UNI-2}}      \\ 
	\cline{2-7}
	\begin{tabular}[c]{@{}c@{}}\textbf{Active }\\\textbf{Learning}\\\textbf{Budget}\end{tabular} & \textbf{Mirai} & \textbf{Gafgyt} & \textbf{DEM} & \textbf{Mirai} & \textbf{Gafgyt} & \textbf{DEM}  \\ 
	\hline
	0                                                                                            & 0.19           & 0.13            & 0.21         & 0.18           & 0.06            & 0.22          \\
	10                                                                                           & 0.48           & 0.38            & 0.59         & 0.40           & 0.28            & 0.38          \\
	100                                                                                          & 0.70           & 0.60            & 0.66         & 0.63           & 0.49            & 0.52          \\
	200                                                                                          & 0.79           & 0.73            & 0.69         & 0.76           & 0.69            & 0.59          \\
	300                                                                                          & 0.83           & 0.76            & 0.66         & 0.81           & 0.73            & 0.61          \\
	400                                                                                          & 0.84           & 0.77            & 0.68         & 0.83           & 0.75            & 0.68          \\
	500                                                                                          & 0.87           & 0.81            & 0.70         & 0.84           & 0.76            & 0.69          \\
	Fully Labeled                                                                                     & 0.97           & 0.96            & 0.82         & 0.96           & 0.93            & 0.72         
\end{tabular}

\caption{Active federated learning at different budgets from 0 to 500 in a challenging scenario in which none of the malware families is labeled in training. We show PR-AUC for detecting each malware family at each university networks. The federated model without active learning (budget 0) cannot identify new malware (top row). The active learning component achieves higher PR-AUC at larger budgets, and gets close to the model trained with all labeled malicious data (last row). }

\label{tab:active_federated_table}
\end{table}

We showed how federated learning can transfer knowledge about malware observed in some networks to detect attacks seen for the first time in other networks.
However, the global model will not detect a completely new attack that has not been observed in either of the participating clients in training.  To overcome this challenge, we propose extending the federated learning framework with an active learning component that integrates a local anomaly detection module (Section~\ref{label:activeFL}).

We design a challenging scenario to test the detection capabilities of active federated learning. The two university networks start training the global model using NIDS alerts known to be malicious, but without using the malicious labels to simulate a situation when clients are infected with unknown malware. Thus, we merge the \emph{unlabeled} malware logs from the three malware families (Mirai, Gafgyt, and DEM) into the client networks. 
We would like to test if the anomaly detection module identifies some of the malicious samples in the local networks, and, furthermore, if the federated model with active learning is able to detect these unknown malware attacks when no labeled data from these families is initially available in training. 

For the anomaly detection module, we train Isolation Forest models on HTTP logs, on $k$ previous time windows of 30 minutes each, and use an ensemble of these models to detect anomalies within the current time window. 
We set $k=3$, after experimenting with multiple values of $k$.
We vary the budget used in investigation between 0 (which results in the federated model without active learning) and 500.

Table~\ref{tab:active_federated_table} shows the PR-AUC of detecting each of the three malware families as a function of the investigation budget. As expected, for a budget of 0, the global model performs poorly without active learning at detecting these completely new malware families (e.g., PR-AUC 0.19 at detecting Mirai at UNI-1). 
Active learning used in federated training significantly helps detect new malware: as we increase the budget for investigation, the PR-AUC also increases, and with a budget of 500 samples it reaches PR-AUC of 0.87 in the same experiment for detecting Mirai.

We also trained the active learning system with sample selection using only the anomaly detection module (without the classifier). This setup identified fewer samples per round (since it did not incorporate any samples selected by the classifier), and the detection performance was slightly worse. For instance, using both anomaly detection and the highly ranked samples generated by the classifier resulted in an increase of 11\% PR-AUC compared to using only the anomaly detection module for detecting DEM. This result shows that both components for sample selection contribute to the success of active federated learning.

%% file: evaluation/eval_poisoning.tex
\subsection{Resiliency Against Poisoning}
\label{sec:poisoning}

\begin{figure}[t]
	\centering
	\begin{subfigure}[b]{0.49\linewidth}
		\includegraphics[width=\linewidth]{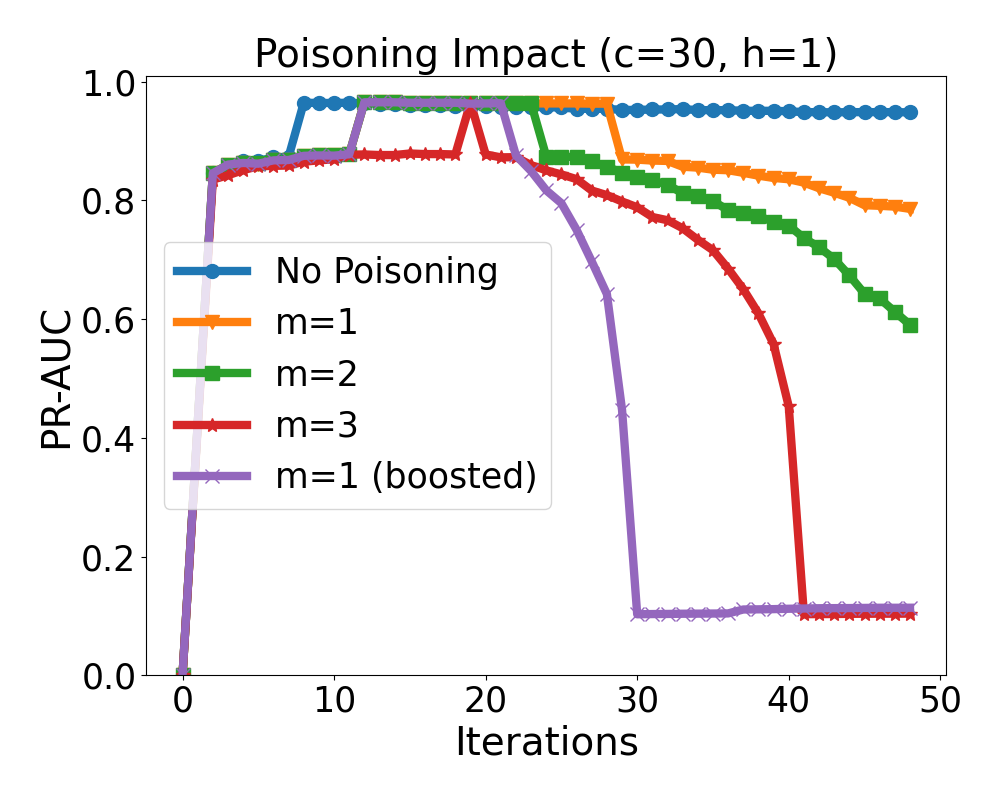}
	\end{subfigure}
	\begin{subfigure}[b]{0.49\linewidth}
		\includegraphics[width=\linewidth]{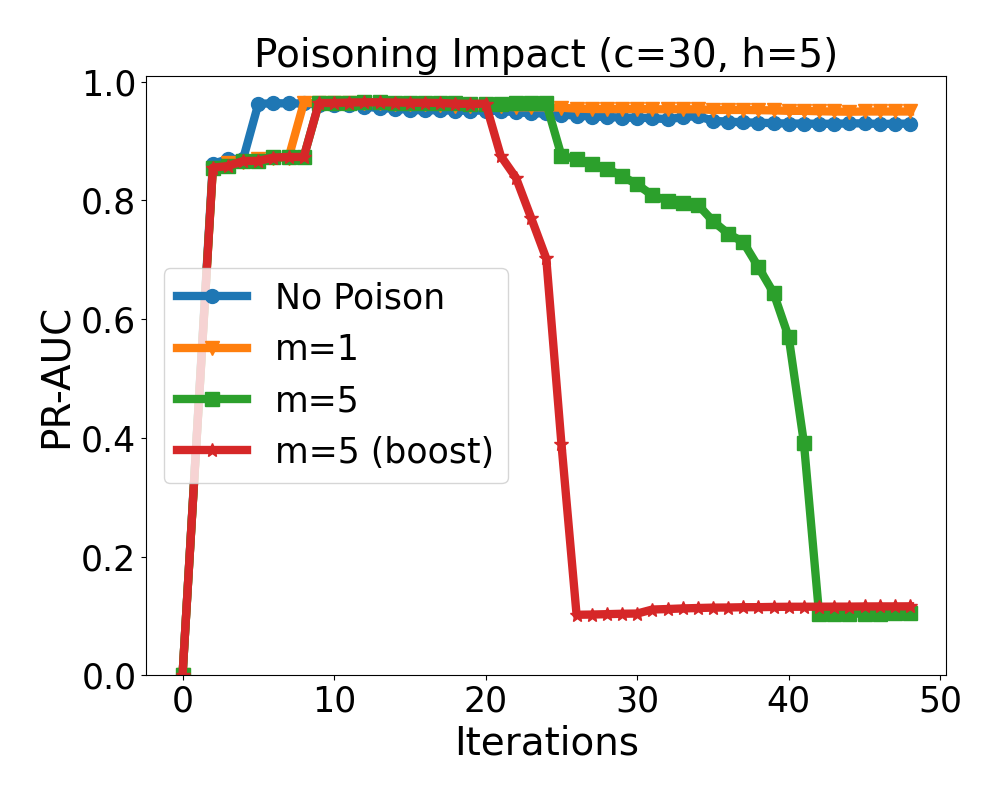}
	\end{subfigure}
	\caption{Poisoning Attacks: Global model performance progress with various attack strategies, and no defense. We compare the `no poisoning' baseline, the label flipping attack with $m$ malicious clients, and a stronger `weight boosting' attack. Poisoning starts at iteration $20$. Increasing the number of helper clients $h$ improves detection: $h = 1$ (left) and $h=5$ (right).}
	\label{fig:pois_attack} 
	\vspace{-10pt}
\end{figure}

\begin{figure}[t]
	\centering
	\begin{subfigure}[b]{0.49\linewidth}
		\includegraphics[width=\linewidth]{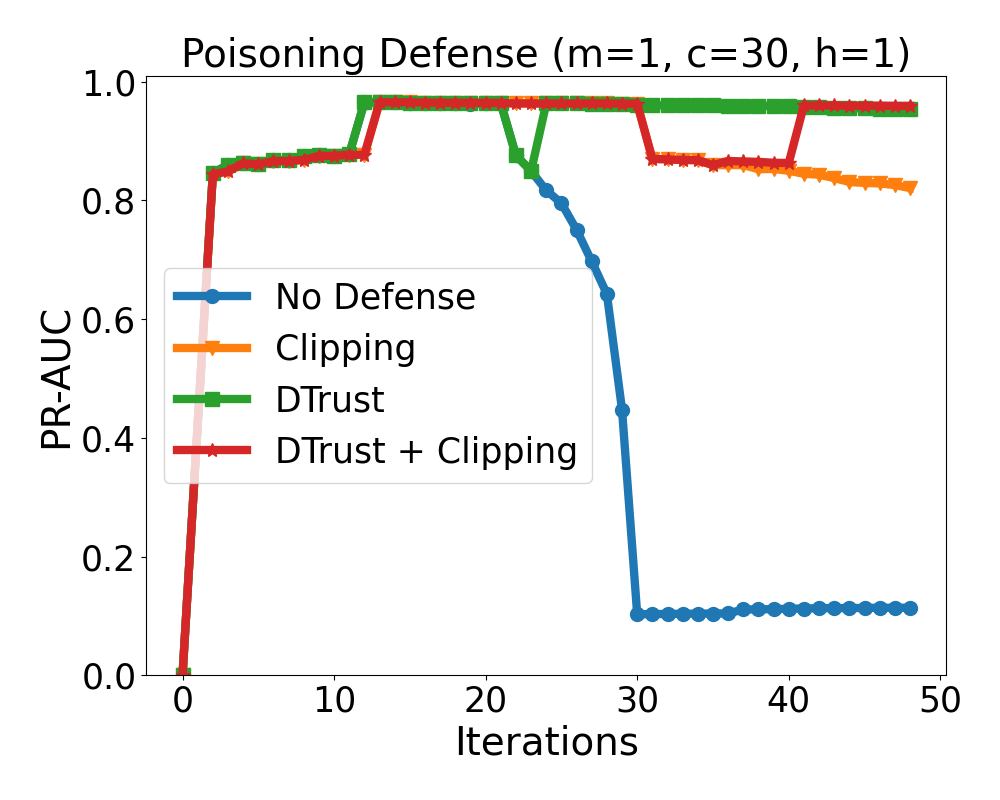}
	\end{subfigure}
	\begin{subfigure}[b]{0.49\linewidth}
		\includegraphics[width=\linewidth]{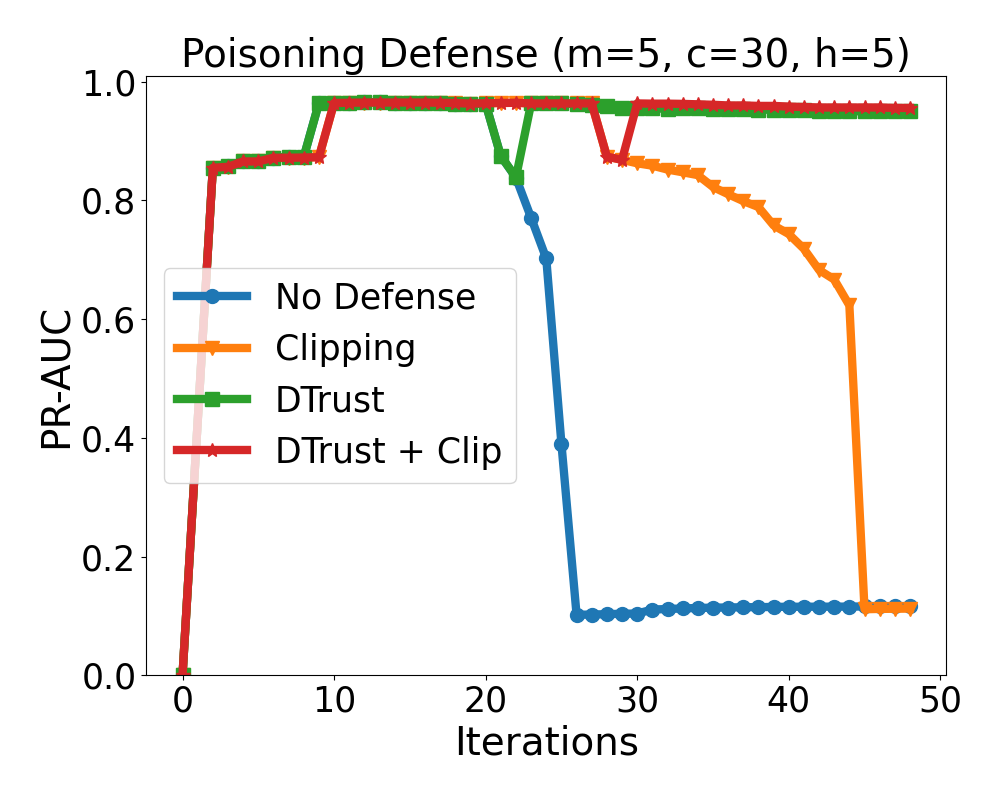}
	\end{subfigure}
	\caption{Defenses: Global model performance progress when various defenses have been employed against the strongest attack from Figure~\ref{fig:pois_attack} (weight boosting).
	DTrust successfully recovers from the poisoning attack, outperforming the `no defense' baseline and the `clipping' defense significantly.}
	\label{fig:pois_defense} 
	\vspace{-10pt}
\end{figure}

In this section, we evaluate \thesystem's resiliency against poisoning attacks, where a number of malicious clients try to impact the global model to evade detection of a specific malware pattern. In our threat model, clients belong to one of the following three categories: (1) \emph{poisoning clients} that try to inject a specific malicious pattern; (2) \emph{helper clients}, which have observed this specific malicious behavior before; their dataset contains correctly labeled data related to the malicious patterns; (3) other benign clients that have not seen the specific poisoning activity before. The helper clients are instrumental in sharing their knowledge through the global model to increase resilience against adversarial actions.

In these experiments, we consider a total of $c = 30$ clients, and vary the number $m$ of malicious clients, as well as the number $h$ of helper clients. We run the training for 48 iterations (one day of data, with 30-min long iterations), and start the poisoning activity at the 20th iteration. We study three poisoning scenarios with specific HTTP attacks: (1) Data exfiltration activity to the C2 server from CosmicDuke (part of our DEM dataset)~\cite{cosmicDukeAnalysis}; (2) ThinkPHP exploit observed in Mirai~\cite{ThinkPHPAttack}; and (3) a home-router attack with device update behavior from Gafgyt~\cite{huaweiRouterAttack}.

In Figure~\ref{fig:pois_attack}, we show the impact of poisoning during the data exfiltration attack, when no defense is employed. We compare the `no poisoning' baseline with: (1) a label flipping attack, where a number of malicious clients $m$ provide corrupt updates trained on malicious data labeled as benign; (2) a stronger poisoning attack using weight boosting~\cite{bhagoji2019analyzing}, which directly scales up the weight updates sent by the poisoning clients to the server. We make the following observations:
First, we note that the success of the label flipping attack increases with more poisoning clients $m$, as expected. 
Second, the global model is more resilient against poisoning attacks when more helper clients are contributing with locally seen attack patterns. With only one helper, a single malicious client is sufficient to decrease model accuracy over time substantially (left figure, with boosting); however, when five helpers participate in the global model, the attacker needs to compromise as many as five clients for an equivalent performance degradation (right figure, with boosting).
Third, we note that weight boosting attacks are more effective than label flipping attacks, reaching PR-AUC of $0.11$ in both experimental settings (i.e., left and right figures) in only 10 and 6 iterations (after poisoning had started), respectively.

In Figure~\ref{fig:pois_defense}, we explore possible defenses against the weight boosting attack presented before. We compare the `no defense' baseline with: (1) the weight clipping technique~\cite{sun2019can}; (2) DTrust algorithm; and (3) a hybrid defense consisting of DTrust + clipping. We used clean training updates to determine the clipping bound parameter as $0.1$, after observing that lower values hinder the main task accuracy significantly. 
The plot on the right ($m = 5, h = 5$) demonstrates that the clipping technique alone is not sufficient to mitigate the impact of poisoning, as the PR-AUC still drops to $0.11$ eventually during this strong attack.
Our proposed defense, DTrust, starts to investigate the client updates after the global model PR-AUC deteriorates by more than 10\%. DTrust identifies the malicious clients and removes their updates from the global model. 
The training continues with the benign clients, and the performance recovers over time. Eventually, DTrust reaches $0.93$ PR-AUC, similar to the `no poisoning' case.  DTrust can be used as a standalone defense, as well as in combination with clipping (giving similar results). 
CELEST is also effective in other poisoning scenarios with different HTTP attacks.  Table~\ref{tab:pois_scenarios} summarizes the poisoning impact for the three attacks after one day of training, and presents the mitigation results. DTrust detects the poisoning activity in less than 4 iterations in each case. Furthermore, DTrust enables poisoned models to recover and reach close to the accuracy of the models trained on clean data.

\begin{table}[t]
	\centering
\begin{tabular}{cccc|c}
	\begin{tabular}[c]{@{}c@{}}\textbf{Training}\\\textbf{Family}\end{tabular} & \begin{tabular}[c]{@{}c@{}}\textbf{Poisoning}\\\textbf{Behavior}\end{tabular}  & \begin{tabular}[c]{@{}c@{}}\textbf{No}\\\textbf{Defense}\end{tabular} & \begin{tabular}[c]{@{}c@{}}\textbf{DTrust }\\\textbf{Defense}\end{tabular}& \begin{tabular}[c]{@{}c@{}}\textbf{Clean}\\\textbf{Training}\end{tabular}\\
	\hline
	Mirai  & Data Exfiltration    & 0.11  & \textbf{0.93}  & 0.93 \\
	DEM & HomeRouter Attack       & 0.08  & \textbf{0.89}  & 0.92 \\
	DEM  & ThinkPHP Exploit       & 0.06  & \textbf{0.65}  & 0.70                        
\end{tabular}
	\caption{Overview of the poisoning experiments for three scenarios after one day of training in a 30-client setting. 
	The training family is the main task that all clients train on. We run the weight boosting attack with $m = 5$ malicious clients, and $h = 5$ helpers. 
	Our DTrust defense reaches PR-AUC scores close to the clean model by detecting the poisoning activity in less than 4 iterations in all three cases.}
	\label{tab:pois_scenarios}
\end{table}

%% file: evaluation/eval_deployment.tex
\section{Model Deployment}
\label{sec:deployment}

\begin{figure}[t]
	\centering
	
			\includegraphics[width=0.9\linewidth]{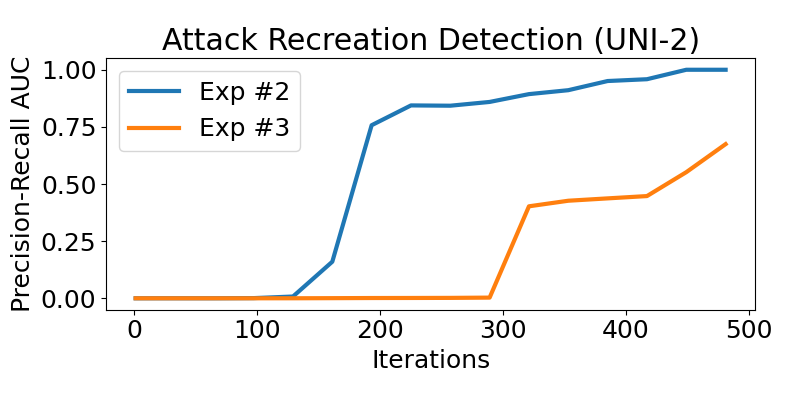}
    	\vspace{-8pt}


	\caption{PR-AUC of the global model trained at various time intervals. Training data includes labeled attack recreation samples used at UNI-1, but not at UNI-2, in Exp 1. Testing is done on  Exp 2 and Exp 3 at UNI-2.  Federated model reaches almost perfect PR-AUC in Exp 2, after the model is trained with sufficient data, due to the similarity of Exp 1 and Exp 2. In Exp 3, the PR-AUC is lower and the model needs more training data, due to adversarial evasion. }	 
	\label{fig:attack_rec_iters} 
\end{figure}

\begin{table}[t]
\centering
\begin{tabular}{c|ccc}
\multirow{2}{*}{\begin{tabular}[c]{@{}c@{}}\textbf{Training }\\\textbf{ Time (Days)}\end{tabular}} & \multicolumn{3}{c}{\textbf{Detection on UNI-2}} \\ 
\cline{2-4}
 &\textbf{Exp \#1} &\textbf{Exp \#2} &\textbf{Exp \#3}\\ 
\hline
3  & 0.2449  & 0.1601 & 0.0006  \\
6  & 0.8258  & 0.8592  & 0.4025  \\
10 & 0.9272  & 1.0     & 0.6746            
\end{tabular}
\caption{PR-AUC at UNI-2 during the 3 attack recreation experiments for the global model trained with 3, 6, and 10 days of Exp 1 data.  Training data includes labeled attack recreation samples used at UNI-1, but not at UNI-2. The model trained after 3 days has relative low PR-AUC, but this increases as more training data is used. Results on Exp 1 and Exp 2 have higher PR-AUC than Exp 3, which employs  evasion.}
\label{tab:deployment_stats}
\end{table}

\begin{table}[t]
\centering
\begin{tabular}{c|cccccc}
\textbf{}                                                                  & \multicolumn{6}{c}{\textbf{Deployed Network: UNI-1}}                                         \\ 
\hline
\begin{tabular}[c]{@{}c@{}}\end{tabular} & \textbf{ARD} & \textbf{IDS} & \textbf{VT} & \textbf{Unk} & \textbf{Prec} & \textbf{FPR}      \\ 
\hline
\textbf{Top-10}                                                            & 8            & 2            & 0           & 0            & 1.0           & 0                 \\
\textbf{\textbf{Top-}30}                                                   & 28           & 2            & 0           & 0            & 1.0           & 0                 \\
\textbf{\textbf{Top-}50}                                                   & 39           & 7            & 2           & 2            & 0.96          & $2.7\times10^-6$  \\
\textbf{\textbf{Top-}80}                                                   & 46           & 19           & 10          & 5            & 0.93          & $6.9\times10^-6$  \\
\textbf{\textbf{Top-}100}                                                  & 59           & 19           & 12          & 10           & 0.90          & $1.3\times10^-5$ 
\\
\hline
\textbf{}                                                                  & \multicolumn{6}{c}{\textbf{Deployed Network: UNI-2}}                                         \\ 
\hline
\textbf{Top-10}                                                            & 9            & 1            & 0           & 0            & 1.0           & 0                 \\
\textbf{\textbf{Top-}30}                                                   & 28           & 2            & 0           & 0            & 1.0           & 0                 \\
\textbf{\textbf{Top-}50}                                                   & 31           & 12           & 1           & 6            & 0.98          & $1.4\times10^-5$  \\
\textbf{\textbf{Top-}80}                                                   & 42           & 14           & 16          & 9            & 0.88          & $2.1\times10^-5$  \\
\textbf{\textbf{Top-}100}                                                  & 42           & 14           & 30          & 14           & 0.86          & $3.3\times10^-5$ 
\end{tabular}


\caption{Detection statistics for global model deployment in the two networks on a  test day part of Exp 1 (June 20). The majority of detected samples are attack recreation (ARD) logs (59 out of 100 at UNI-1), and several samples are confirmed malicious by IDS (19 at UNI-1 and 14 at UNI-2). The system detected 42 new malicious samples with unique URLs (a total of 20 malicious domains), confirmed by VirusTotal (VT), which are not part of ARD and not detected by IDS. The model detects all ARD samples with high precision and low false positive rates in both networks. (Unk = Unknown)}
\label{tab:ranking_stats}
\end{table}

In this section, we discuss results of our system deployment on two university networks to demonstrate its performance and feasibility in a real-world setting. We design several attack recreation exercises to run real malware on the two networks in a controlled manner, and then evaluate the detection capabilities of \thesystem.

\vspace{2pt}
 \BfPara{Attack Recreation Exercises}
We deploy a set of vulnerable virtual machines to recreate a malware infection and propagation on the two networks. We use a modified instance of Mirai~\cite{antonakakis2017understanding}, which scans and propagates to other vulnerable virtual machines under our control. To ensure there were no unintended side effects or damage to other nodes on the internet, the Mirai malware was modified  by updating the dictionary used to brute-force passwords to include only randomly selected user names and passwords.  Mirai's infection component was enhanced to verify that the IP to infect was on the controlled IP list before malware propagates and infects the target host.
We also host a C2 server on the Chameleon~\cite{keahey2020lessons} service outside the university networks. This deployment allows the network data to be collected at the edge of the university networks and we observe the C2 communication in the HTTP logs.  We further enhanced Mirai to deploy a reverse shell component, called Sandcat, from  Mitre Caldera~\cite{sandcat}, to maintain  a permanent C2 connection with beaconing between 700 and 900 seconds on port 2323.

We start an experiment by infecting a few of the machines with the Mirai malware. The lateral movement process scans public IPs across the two universities, and takes approximately two weeks to infect approximately half of the vulnerable nodes. We run three exercises with different configurations, each over a two-week interval:

\begin{itemize}[nosep]
	\item {\bf Exp 1}: A total of 282 end points were deployed, and the C2 server IP was configured to change every 30 minutes. The new server IP was detected by the infected end points by querying a DNS server for a single domain name.  This experiment occurred in June of 2021.
	\item {\bf Exp 2}: Similar to Exp 1, but with 429 end points deployed.  This experiment occurred in August of 2021.
	\item {\bf Exp 3}: Both the C2 server IP and domain change every 30 hours. The new domain name is found by using a domain generation algorithm (DGA) configured to psuedo-randomly select 100 domains from 10,000 possible domain names. 
	This experiment used 364 end points, and occurred in October of 2021.
\end{itemize}

Exp 2 is designed to be similar to Exp 1, but is run after two months to test if the global model maintains its detection performance over time. Exp 3 is designed to emulate an adversarial evasion strategy, in which the adversary rotates both its C2 domain and IP infrastructure to evade detection.

\vspace{2pt}
 \BfPara{Deployment and Evaluation}
 We deploy the federated learning framework in the two networks, and perform  training over ten days in June 2021, starting two days before Exp 1 began to account for some intervals without any attack recreation samples. Commercial off-the-shelf network IDS (NIDS) logs which are commonly available as part of the organization's security infrastructure are used to label traffic in both networks during training.  In this scenario, we have two client organizations, and one of the networks (UNI-1) has resources to detect and label the attack recreation data as malicious. The second network  (UNI-2) has no attack recreation labels during training. These two networks participate in federated training, and try to detect the various attack recreation experiments.

 We are interested in measuring global model detection at UNI-2, for which the local model has 0 PR-AUC (as there is no labeled attack recreation data). Table~\ref{tab:deployment_stats} shows the PR-AUC results at UNI-2 for three models trained on Exp 1, using 3, 6, and 10 days of training data, respectively.  Figure~\ref{fig:attack_rec_iters} shows the increase in PR-AUC on Exp 2 and Exp 3 testing  days (August 15 and October 15) for the same model trained on Exp 1 data, at various time intervals. Initially, there is no infected host at UNI-1 (we start the training 2 days before Exp 1), so the model has no malicious detections until the third day. In the third day, the model has relatively low PR-AUC (0.24 on Exp 1 testing), but the performance increases steadily with more training data (and reaches a perfect PR-AUC on Exp 2 after 10 days). Interestingly, the global model trained on Exp 1 generalizes very well on Exp 2, which occurs two months later. This outcome is explained by the similarity in the attack recreation data in the two experiments,  which has a higher relevance than the changes in the benign data distribution over time. Results on Exp 3 are surprisingly good (PR-AUC of 0.67), considering the evasion performed by malware, which changes both the C2 server domain and IP address. These results suggest that, while IP and domain features are important, \thesystem\ relies on a larger number of features extracted from HTTP logs and provides some resilience to evasion.

We further analyze the highest ranked HTTP logs on the test day, June 20, in Exp 1 to determine what the model considers suspicious in addition to attack recreation events. 
Table~\ref{tab:ranking_stats} includes the results of our investigation of top ranked samples in both networks. We observe that most of the detections are attack recreation samples (28 in top 30, and 46 in top 80 samples), and a large number (19 on UNI-1 and 14 on UNI-2) are verified as malicious by the commercial IDS. We query the remaining samples on VirusTotal and determine 12 samples at UNI-1 and 30  at UNI-2 with positive scores (a total of 42 malicious URLs and 20 malicious domains). These detections are interesting as they are new malicious detections  identified by \thesystem, which were not part of the attack recreation exercises and the IDS failed to detect. The rest (10 URLs at UNI-1 and 14 at UNI-2) have an unknown status at the time of the investigation. We report high precision and  FPR lower than $3.3\times10^{-5}$ in both networks, considering the unknown samples as false positives.

We also deployed our model for real-time  detection during attack recreation experiments. We ran a network monitoring service that consumes the HTTP log streams from the live network traffic.
Features are generated in real-time from raw log samples, and evaluated on the global model to generate a score.
For high-certainty detections with scores above 0.85, we trigger a firewall rule to prevent future communications with the detected C2 server. Our model was able to detect the C2 malicious communications, and deployed the firewall successfully to block the attack in progression.

%% file: discussion.tex
\section{Discussion and Extensions}

In this paper we showed that global models that leverage coordinated detection across multiple participating organizations perform better at detecting malware threats compared to local models that only rely on limited local data. Federated learning is often able to expose new
attacks which are largely invisible to individual organizations. FL is particularly effective when used in conjunction with active learning, which progressively enhances the labeled dataset used in training with new malicious samples. When deployed on two university networks, \thesystem\ was able to detect new malicious activity and identify evasive adversarial activity in real time. 

\vspace{2pt}
\BfPara{Real-world Deployment} 
Active learning adds a major boost to the malware detection capabilities of a federated learning framework. Through the anomaly detection module integrated with the classifier that progressively improves its detection, \thesystem\ can discover and learn the behavior of completely new attacks for which no labeled data is available. Our current framework uses a generic anomaly detector (Isolation Forest); however, in order to tap into its full potential, a more sophisticated anomaly detector specifically designed for HTTP logs can be used. 
We showed in the evaluation that active learning can be effective even at small budgets of samples investigated and labeled by security analysts.
We note that periodic training is necessary in order to learn changing patterns in background traffic, as well as to discover and incorporate new attack behaviors in the model. 
While \thesystem\ is currently designed to detect malware communication from HTTP logs, incorporating other types of security logs (e.g., authentication logs, DNS logs, firewalls logs) would be beneficial. \thesystem\ would have less information if the malware traffic is encrypted over HTTPS, but we have shown that it still performs well when only a subset of features are available (see Appendix~\ref{apx:exp_features}). 
An interesting extension is to support P2P architectures, where a group of autonomous peers jointly train a common model without using a trusted server. A decentralized FL approach avoids having a single point-of-failure and several designs have recently been proposed~\cite{Chou2021EfficientAL,Gholami2022ServerlessFL,wink2021p2pfl}. 

\vspace{1pt}
\BfPara{Resilience to Poisoning}
Other studies have proposed to use a trust dataset~\cite{DBLP:conf/ndss/CaoF0G21} which resides on the server and is used to remove outlier updates or correct the model. In contrast, we employ a distributed trust bootstrapping approach where each client has its own trust dataset and actively participates in defense. 
We note that federated learning with privacy-preserving aggregation does not allow inspection of individual updates from the clients~\cite{bonawitz2017practical}. Our defense can be modified to simply revert to a previous model, but identification of malicious clients would not be possible in this case.

\vspace{1pt}
\BfPara{Resilience to Evasion} 
In Section~\ref{sec:deployment} we presented a case study (Exp 3) designed to emulate an adversarial evasion strategy, in which the adversary rotates both its C2 domain and IP infrastructure to evade detection. We showed that \thesystem\ is still able to detect the malware attack when faced with this evasion method. \thesystem's resilience can be attributed to two main factors: the large number of features used for training the federated model, and the participation of multiple clients.  Still, a motivated attacker might be able to coordinate its behavior and generate traffic that looks normal in all participating clients to evade \thesystem's detection. We believe evasion is more difficult in a federated setting than in a local system trained within a single organization.

%% file: related_work.tex
\section{Related Work}
\vspace{1pt}
\noindent\textbf{Malicious URL and domain detection.}
A large body of work has looked at URL-based detection~\cite{cao2015detecting,le2018urlnet,ma2009beyond,mamun2016detecting,saxe2017expose,yuan2018url2vec,thomas2011design} and domain-based detection~\cite{bilge2014exposure,canali2011prophiler,de2021compromised,li2019machine,zhauniarovich2018survey} of malicious traffic. Some of these studies focus on one type of malware like phishing~\cite{yuan2018url2vec} or spam~\cite{cao2015detecting,thomas2011design} in social media contexts, while others attempt to capture a larger set of malware behaviors, e.g, Mamun \etal~\cite{mamun2016detecting} group malicious URLs into categories like spam, phishing, malware, and defacement URLs using public datasets. 
Previous studies employ various techniques to detect malicious URL and domains. These include lexical features generated from bag-of-words representation~\cite{ma2009beyond}, behavioral analysis~\cite{cao2015detecting}, DNS data analysis~\cite{bilge2014exposure, zhauniarovich2018survey}, webpage content analysis combined with some URL and domain properties~\cite{canali2011prophiler,thomas2011design}, and deep-learning methods~\cite{le2018urlnet,li2019machine,mamun2016detecting,saxe2017expose,yuan2018url2vec}.

\vspace{1pt}
\noindent\textbf{Malicious HTTP-based detection.}
Efforts on malware detection in this area have focused on HTTP-based detection using web-proxy logs~\cite{bartos2016,hu2016baywatch,mcgahagan2019comprehensive,nelms2013execscent,oprea2018made,oprea2015detection, stokes2008aladin} and HTTP-based application fingerprinting~\cite{bortolameotti2017decanter, bortolameotti2020headprint,perdisci2010behavioral}.
Machine learning detection methods on HTTP logs attempt to detect malware activity within a single network by training supervised classification models locally on labeled data available in that network~\cite{bartos2016,hu2016baywatch,oprea2018made,oprea2015detection}. In other approaches, Nelms \etal~\cite{nelms2013execscent} use control protocol templates derived from labeled samples to detect new C2 domain names, while Bortolameotti \etal~\cite{bortolameotti2017decanter} use application fingerprinting techniques for clustering HTTP connections in order to detect anomalous traffic. 
 
\vspace{1pt}
\noindent\textbf{Federated learning for cyber security.}
Federated learning has been proposed for enhancing cyber security in various settings including mobile phones~\cite{galvez2021less}, Internet-Of-Things~\cite{nguyen2019diot, schneble2019attack}, cloud ecosystems~\cite{payne2019towards}.
Closer to our work, Khramtsova \etal~\cite{khramtsova2020federated} study federated learning approaches on malicious URL detection to show the benefit of sharing information about local detections. Zhao \etal~\cite{zhao2019multi} propose multi-task network anomaly detection using federated learning. Fereidooni \etal~\cite{fereidooni2022fedCRI} proposed federated learning to enable effective cyber-risk intelligence sharing for mobile devices. 
Several studies have looked at poisoning attacks against federated learning~\cite{bagdasaryan2020backdoor,sun2019can,tolpegin2020poisoning,fang2020local,Xie2020DBA,shejwalkar2021manipulating,Wang21_Tails}.
Fang et al.~\cite{fang2020local} proposed a general framework of local model poisoning attacks, which can be applied to optimize the attacks for any given aggregation rule. Bagdasaryan et al.~\cite{bagdasaryan2020backdoor} has formulated adversarial poisoning as a two-task optimization problem that has high accuracy on both the main and the backdoor tasks (constrain-and-scale method). Previously proposed defenses against poisoning in FL perform anomaly detection on client updates~\cite{blanchardMachineLearningAdversaries2017,mhamdiHiddenVulnerabilityDistributed2018,yinByzantineRobustDistributedLearning2018a}, are specific to backdoor attacks~\cite{rieger2022deepsight,nguyen2022flame}, or leverage a trusted dataset at the server~\cite{DBLP:conf/ndss/CaoF0G21}.

%% file: conclusion.tex
\section{Conclusion}

In this paper, we present \thesystem, a federated learning framework for collaborative threat detection. \thesystem\ leverages a distributed machine learning architecture in which multiple participating organizations train a global model used for HTTP-based malware detection. 
Using a novel active learning component, \thesystem\ progressively improves its detection capabilities.
In addition, we propose DTrust, a new resilient algorithm aimed at defending against data and model poisoning attacks in distributed settings.
We evaluate our system using a variety of malware datasets and demonstrate the power of knowledge transfer through the globally trained model, which enables individual organizations to detect attacks that were largely invisible locally. We deploy the model on two large university networks and show that \thesystem\ is able to detect real-world malicious traffic (42 malicious URLs and 20 malicious domains). Overall, \thesystem\ is a scalable and effective proactive threat detection solution that leverages collaboration across multiple networks to detect emerging cyber threats.

%% file: appendix.tex
\appendix

\input{appendix/shortened_features}

%% file: appendix/shortened_features.tex
\section{Feature Extraction from HTTP Logs}
\label{apx:features}

\begin{table}[thpb]
	\small
	\centering
	
	\scalebox{0.85}{
	\begin{tabular}[width=1\textwidth]{|>{\raggedright\arraybackslash}m{0.6in} |>{\raggedright\arraybackslash}m{0.9in} |>{\raggedright\arraybackslash}m{1.3in} |}
		\hline 
		\bf{Type} & \bf{Fields} & \bf{Description} \\
		\hline \hline
		{\bf{Embedded features}} & URL, Domain Name and Referer  & Based on  embedding models that preserve URL and domain structure. (2176 features)\\ \hline
		{\bf{Numerical features}} & Request and Response Size, Transaction Depth, Version & Represented as float. (4 features)\\\cline{2-3}
		& UA Browser Information & Browser version  from UA represented as float. (2 features) \\ \cline{2-3}
		& Binary Features for host, URI, referer, user agent, method & Binary features to indicate the existence of certain fields in the HTTP log. (5 features)\\ \hline
		{\bf{Categorical features}} & External IP   & IP subnets are represented as a sequence of three tokens corresponding to the octets, and then one-hot-encoded. (768 features) \\ \cline{2-3}
		& Port &  Destination port, one-hot-encoded. (100 features) \\ \cline{2-3}
		& User Agent  &User agents are parsed to retrieve device, browser, and OS, which are one-hot encoded. (2542 features)\\
		\cline{2-3}
		& HTTP Method, Status Code, Content Type  & Represented as one-hot-encoded vectors. (267 features) \\
		\hline
	\end{tabular}
	}
\caption{Features extracted from HTTP logs. Each HTTP log record is represented as a fixed-size vector containing numerical representations of these features.}
\label{tab:features}
\end{table}

Table~\ref{tab:features} presents the features used in this study. 

\BfPara{Embedding model design} We considered several approaches for generating URL embeddings. 
The first option is a centralized training of the word models using either Word2Vec or FastText architectures.  The training of the embedding model is carried out on the server using the data collected from all the clients. Thus, a single entity (i.e., the server) has access to the entire data, and uses it to learn the vector embeddings from the URLs (or domain, referer). This approach suffers from scalability and privacy issues, as all the raw URL data has to be collected in a central server to perform the training.

We are interested in a more privacy-preserving and scalable approach for distributed training of the embeddings among multiple participants. In theory, it is possible to apply the Federated Averaging algorithm to train Word2Vec or FastText models.  However, in practice, there are a number of challenges that prevented us for adopting this solution. First, existing libraries for word embedding generation  such as GenSim are highly optimized (e.g., they have custom C implementations and parallelize gradient updates using multiple cores), and it is not immediately clear how to leverage these libraries in combination with the Federated Averaging algorithm (which requires server-side aggregation of local model updates). Second, Federated Averaging for Word2Vec has been shown to exhibit slow convergence due to the large size of updates sent in each iteration~\cite{bernal2021federated}.

Based on these considerations, we developed a distributed approach using integrated sequential client updates. In this approach, each client has access to its own data and updates the global embedding model locally, using its entire corpus. The client sends the updated global model back to the server, who acts as a trusted central coordinator. The clients apply their updates sequentially, in a round robin fashion, over multiple iterations.

When instantiated with Word2Vec, this method still requires a  common word (i.e., token) vocabulary. To address this requirement, the clients send their token frequencies to the server in a pre-processing phase. The server aggregates them, deciding on a lower bound frequency (e.g., all tokens that appear more than once), and sends the global vocabulary back to each client. The privacy of the system can be further enhanced with FastText embeddings, which use n-grams (i.e., sequences of n characters) as tokens. This method will alleviate the need to share tokens in advance, as we can consider all the possible n-grams to constitute the vocabulary. FastText has the additional advantage of supporting new tokens not observed at training time by generating n-gram representations for them. This property is an important one, as adversaries can change parts of the URLs to evade detection (for example, the query string, path, and parameters can be easily updated). As we show in our evaluation, FastText and Word2Vec perform similarly in performance (and have higher performance than lexical features, as expected). For these reasons, we select federated FastText as the preferred embedding method for \thesystem\ URL representation.